\documentclass[twocolumn]{aastex631}
\usepackage{lmodern}

\usepackage{newtxtext,newtxmath}
\usepackage[T1]{fontenc}

\usepackage{graphicx}	
\usepackage{amsmath}	
\let\longtable*\relax
\usepackage{enumerate}

\usepackage{color}
\usepackage{soul}

\newcommand{\hei}{He\,\textsc{i}\,$\lambda$5876}
\newcommand{\heii}{He\,\textsc{ii}\,$\lambda$4686}
\newcommand{\ha}{H$\alpha$}
\newcommand{\hb}{H$\beta$}
\newcommand{\dashtwo}{NGC 1624-2}

\defcitealias{2012Wade.NGC.1624.2}{W12}
\defcitealias{2021MNRAS.501.2677D}{DU2021}
\defcitealias{2021MNRAS.501.4534J}{J2021}

\accepted{February 9, 2026}

\submitjournal{ApJ}

\begin{document}

\title{New Insights from Revisiting the Rotation Period of the Strongly Magnetic O Star, NGC 1624-2}

\correspondingauthor{Shaquann Seadrow}
\email{sseadrow@udel.edu}
\author{S. Seadrow}
\affiliation{Department of Physics \& Astronomy, Bartol Research Institute, University of Delaware, Newark, DE, 19714, USA}

\author{V. Petit}
\affiliation{Department of Physics \& Astronomy, Bartol Research Institute, University of Delaware, Newark, DE, 19714, USA}

\author{G. A. Wade }
\affiliation{Department of Physics, Engineering Physics \& Astronomy, Queen's University, Kingston, Ontario K7L 3N6, Canada}
\affiliation{Department of Physics \& Space Science, Royal Military College of Canada, PO Box 17000, Station Forces, Kingston, ON K7K 7B4, Canada}

\author{D. Bohlender}
\affiliation{Dominion Astrophysical Observatory, Herzberg Astronomy and Astrophysics Program, National Research Council of Canada \\ 5071 West Saanich Road, Victoria, BC V9E 2E7, Canada}

\author{J. Ma{\'i}z Apell{\'a}niz}
\affiliation{Centro de Astrobiolog{\'i}a (CAB), CSIC-INTA, Campus ESAC. C. Bajo Del Castillo S/N, E-28 692 Madrid, Spain }

\author{A. David-Uraz}
\affiliation{Department of Physics, Central Michigan University, Mount Pleasant, MI, 48859, USA}

\author{M. Oksala}
\affiliation{Department of Physics, California Lutheran University , Thousand Oaks, CA,  91360, USA}
\affiliation{LESIA, Paris Observatory, PSL University, CNRS, Sorbonne University, Universit{\'e} Paris Cit{\'e}, 5 place Jules Janssen, Meudon, France}

\author{J. MacDonald}
\affiliation{Department of Physics \& Astronomy, Bartol Research Institute, University of Delaware, Newark, DE, 19714, USA}



\begin{abstract}

NGC 1624-2 hosts the strongest surface magnetic field found on an O star thus far. When applied across several epochs of observations, the star's currently accepted rotation period (157.99 d) does not coherently characterize the variations of spectral lines of magnetospheric origin. We analyze Lomb-Scargle periodograms produced with new and archival, multi-instrument spectroscopic time series of Balmer H and He spectral lines. We find that 153.17 $\pm$ 0.42 d and 306.56 $\pm$ 1.19 d are both equally suitable periods at phasing the spectral and magnetic time series data in a manner consistent with the Oblique Rotator Model. The 306.56 d period implies a magnetic geometry for NGC 1624-2 that is quite different from the previously accepted one, for which both magnetic poles should be observed during a full rotational cycle. If this is the case, the star's magnetic South pole has yet to be observed, and additional spectropolarimetric observations should be acquired in order to confirm whether or not the south pole is in fact observable.

\end{abstract}



\section{Introduction}
\label{sec:intro}

An estimated  7\% of Galactic O stars possess detectable surface magnetic fields \citep{2017MNRAS.465.2432G}. These fields are organized (predominantly large-scale dipoles), strong (1 to 10 kG), and stable. Currently, the mechanism generating these fields is poorly understood. The leading hypothesis suggests that these fields are of fossil origin: the magnetic fields are either byproducts of main-sequence star mergers or remnants of the Interstellar Medium’s magnetism. \citep[see e.g.][for a recent review]{2023Galax..11...40K}. Their origins notwithstanding, magnetic O stars serve as laboratories to test the dynamical interactions of large-scale, surface magnetic fields with stellar properties such as rotation and mass loss \citep[e.g.][]{2008MNRAS.385...97U,2011A&A...525L..11M,2017MNRAS.466.1052P}.  

NGC 1624-2, a main-sequence O7f?p star, hosts the strongest known magnetic field found on a main-sequence O star. Its magnetic field is measured to be $\sim 20$kG \citep[][hereafter \citetalias{2012Wade.NGC.1624.2}]{2012Wade.NGC.1624.2}, roughly an order of magnitude stronger than the  other magnetic O stars  \footnote{ We specify \textit{main-sequence}   magnetic O stars because  the recently discovered 40 kG, magnetic quasi-Wolf-Rayet star, HD 45166, exhibits O-type spectral features and the rare Of?p peculiarity \citep{2023Sci...381..761S}. However, as a stripped $\sim 2\textrm{M}_{\odot}$ star, its initial magnetic and stellar evolution remains uncertain.}. This star provides the opportunity to explore strong magnetic field effects in the high-mass regime.

In this paper, we present evidence that either i) the rotational period of NGC 1624-2 needs to be revised downwards by a few days, affecting the phasing of recent observations, or ii) the rotational period of NGC 1624-2 is, in fact, twice as long as originally reported, with significant consequences for the geometry of the magnetic field. 

\subsection{Background on the ORM and magnetosphere model}

Rotation periods of most magnetic massive stars can be measured by exploiting the misalignment between the magnetic field and the rotational axis (by an angle $\beta$ known as the \textit{obliquity}). As long as the star's rotation axis is also tilted (by an \textit{inclination} angle $i$) with respect to the line of sight, our view of the surface field changes as the star rotates. Therefore, magnetic field measurements like the mean longitudinal field ($\langle B_\textrm{z} \rangle$) and the mean magnetic modulus ($\langle B \rangle$) will vary periodically \footnote{$\langle B_\textrm{z} \rangle$ is the brightness-weighted longitudinal component of the surface field averaged over the stellar disk. $\langle B \rangle$ is the brightness-weighted magnitude of the surface field averaged over the stellar disk.}. This is referred to as the Oblique Rotator Model \citep[ORM;][]{1950MNRAS.110..395S}. 

 For some stars, the mean modulus can be measured using the  Zeeman splitting that is seen directly in their spectra \citep[e.g.][]{2017A&A...601A..14M}. However, massive stars have other competing line-broadening mechanisms, so they are rarely observed. Instead, the variation of $\langle B_\textrm{z} \rangle$ is  usually studied, but this requires spectropolarimetric observations to measure the change in circular polarization (Stokes $V$) across the line profiles \citep{1997MNRAS.291..658D}. 

Alternatively, some rotation periods of magnetic massive stars can be measured using only spectroscopy. Strong magnetic fields will channel and confine the stars’ ionized, radiatively-driven winds into circumstellar \textit{magnetospheres} with associated emission. These magnetospheric boundaries, characterized by the Alfv\'{e}n radii, can extend multiple times the stellar radius \citep{2002ApJ...576..413U,2008MNRAS.385...97U, 2009MNRAS.392.1022U}.
For stars without dynamically significant rotation, which is common for magnetic O stars, the wind material becomes highly concentrated around the magnetic equator. For tilted magnetic fields, emission from the magnetosphere, such as H$\alpha$, is variable and rotationally modulated \citep{2016MNRAS.462.3830O}. 
Though spectroscopy of magnetospheric emission is more convenient than spectropolarimetry in measuring rotation rates, it bears the disadvantage of not being sensitive to magnetic field polarity.

We often establish a clock around the time of maximum and minimum magnetospheric emission, which are referred to as "high state" and "low state", respectively \citep[e.g.][]{2007MNRAS.375..145N}. A high state in magnetospheric emission occurs when the magnetic axis, \textit{regardless of polarity}, is as close as possible to the line of sight, so that the bulk of the dense magnetospheric material (tracing more or less the magnetic equator) is seen as close to face-on as possible. A high state coincides with an extremum (the non-zero absolute maximum or minimum) of the longitudinal magnetic field. A low state in magnetospheric emission occurs when the bulk of the dense magnetospheric material is seen as close to edge-on as possible; this means that the magnetic equator will be as close as possible to our line of sight. Depending on whether or not the longitudinal field curve crosses zero, the associated longitudinal magnetic field will either be zero or values close to zero. 

For predominantly dipolar magnetic fields, magnetospheric emission cycles exhibit either single-wave or double-wave variation. 
\begin{itemize}
\item Single-wave variation arises from a geometry in which only one magnetic pole (North or South) comes into view over the course of the rotation cycle. We will only see one high state per rotation, and generally, the longitudinal field curve does not cross zero but rather remains consistently positive or negative. At most, the hidden pole will appear at the limb of the stellar disk, in which case the minimum of the longitudinal field curve reaches zero. 
\item Double-wave variation arises from a geometry in which both poles will come into view over the course of the rotation cycle. We will see two (potentially unequal) maxima of emission (high states) per rotation, and the longitudinal field curve crosses zero twice per rotation. One high state will coincide with the positive extremum of the longitudinal magnetic field curve, and one will coincide with the negative extremum; as with their high states, the North and South longitudinal magnetic field measurements can potentially be unequal in magnitude.
\end{itemize} 

However, determining the rotation period solely by magnetospheric emission can lead to ambiguity in the inferred magnetic geometry, such as when double-wave variations have identical high state amplitudes. Thus, magnetic measurements can lift this degeneracy.

\subsection{Previous studies of NGC 1624-2}

The magnetic field of NGC 1624-2 was first detected by \citetalias{2012Wade.NGC.1624.2} using spectropolarimetric observations taken at two epochs. Strong Stokes $V$ signatures were present in multiple spectral lines, and the associated maximum $\langle B_\textrm{z} \rangle$ was 5.4 kG. Additionally, it is the only magnetic O star to have Zeeman splitting observed in its spectra, specifically C\textsc{iv} ${\lambda}5801$ and C\textsc{iv} ${\lambda}5812$. That Zeeman splitting yielded $\langle B \rangle$ $\sim$14 kG (maximum value). With only two measurements to apply the ORM to, their estimated $\sim$20 kG dipolar field strength is a lower limit. 

\citetalias{2012Wade.NGC.1624.2} estimated the extent of the magnetosphere, using the formulation of \citet{2002ApJ...576..413U}, to be around 10$R_\star$ and they indeed noted many rotationally modulated emission lines.
They used a large sample of spectroscopic observations to measure the equivalent widths of these emission lines and inferred a rotational period of 157.99 d that produced a single-wave variation.
We note that as there were only two spectropolarimetric observations at the time, the longitudinal field measurements were not used to determine this period. That said, the two positive $\langle B_\textrm{z} \rangle$ measurements were compatible with the proposed single-wave magnetic geometry.

\citet[][hereafter \citetalias{2021MNRAS.501.2677D}]{2021MNRAS.501.2677D} and \citet[][hereafter \citetalias{2021MNRAS.501.4534J}]{2021MNRAS.501.4534J} conducted spectropolarimetric follow-up studies of \dashtwo\ with the same dataset, though using different methods. Their set of ESPaDOnS (CFHT) spectropolarimetric Stokes $V$ observations spans 3 years ($\sim$8 completed rotations since HJD$_0$) and has even phase coverage when phased with the 157.99 d period of \citetalias{2012Wade.NGC.1624.2}. Both studies evaluate the variations of $\langle B_\textrm{z} \rangle$ and $\langle B \rangle$ time series, but \citetalias{2021MNRAS.501.2677D} used the FWHM of C\textsc{iv}$\lambda$5812 as a proxy for $\langle B \rangle$ while \citetalias{2021MNRAS.501.4534J} derived $\langle B \rangle$ from the splitting of C\textsc{iv}$\lambda$5812. All of the longitudinal field measurements were positive and phased well with the 157.99 period. Therefore, neither study explored alternative periods nor magnetic geometries. 
That said, both studies arrived at the same conclusion that the field potentially has a topology more complex than a single dipole, given that $\langle B \rangle$ unexpectedly reaches a maximum (rather than a minimum) at the same phase at which $\langle B_\textrm{z} \rangle$ reaches its minimum. 

\citetalias{2021MNRAS.501.2677D} conducted a follow-up UV spectroscopic study (to \citet{2019MNRAS.483.2814D}) of \dashtwo\ as well. \citetalias{2021MNRAS.501.2677D} reported that the wind-sensitive resonance lines exhibit three apparent high states within a full rotation. They hypothesized that this could be a consequence of a slight inaccuracy in the 157.99 d rotation period, as the UV observations were obtained at a later epoch than the optical spectra of \citetalias{2012Wade.NGC.1624.2} and their available spectropolarimetric observations. 
Thus in this paper, we will first demonstrate that new spectroscopic observations indeed show that the rotational period needs to be revised.

We will also demonstrate that a longer double-wave period of the order of 300 d is perfectly compatible with all existing magnetic and magnetospheric observations. 
\citetalias{2012Wade.NGC.1624.2} ruled out a double-wave variation of magnetospheric emission lines, but their conclusion assumed that maxima should be unequal \citep[for an examples, see the \ha\ EW variation of $\sigma$ Ori E in Figure 4 of][]{2012MNRAS.419..959O}. Therefore, a double-wave variation with comparable or equal maxima cannot be ruled out, as we will show.
Unfortunately, the possibility of a double-wave variation and a longer $\sim$ 300 d period was never considered in any study thus far; all magnetospheric or spectropolarimetric studies have solely relied on the 157.99 d period.

We present our observations and data reductions in \S\ref{sec:observations}, and demonstrate that new magnetospheric spectroscopic observations do not phase well with the 157.99 d period in \S\ref{sec:need_new_period}. We present our period determination method in \S\ref{sec:methods}. We present our results for single-wave periods in \S\ref{sec:results_sw_periods} and our results for double-wave periods in \S\ref{sec:results_dw_periods}. We discuss the implications of our results on the inferred magnetic geometry (\S\ref{sec:disc_mag_geo}) and the impact of our results on previous studies of \dashtwo\ (\S\ref{sec:past_studies}). We summarize our conclusions in \S\ref{sec:conclusion}.

\section{ Observations}
\label{sec:observations}

We use a collection of optical spectra from spectroscopic and spectropolarimetric observations (the full log of observations can be found in Table \ref{tab:tab_obs_log}). 

 \textit{ESPaDOnS and NARVAL :} We use the Stokes $I$ (unpolarized) spectra of 23 archival optical spectropolarimetric observations obtained with the Echelle SpectroPolarimetric Device for the Observation of Stars (ESPaDOnS) mounted on the Canada-France-Hawaii Telescope (CFHT) and its twin instrument Narval at the T\'{e}lescope Bernard Lyot. These observations were obtained mostly between September 2012 and November 2013, with an additional observation in September 2015. These are high-spectral-resolution observations (R=65,000) that provide a wavelength coverage from $\sim$ 370 nm to 1050 nm.   
 The first two observations were key to the initial magnetic field analysis of NGC 1624-2 by \citetalias{2012Wade.NGC.1624.2}, and the full set of spectropolarimetric observations was used in the recent magnetic geometry analyses of \citetalias{2021MNRAS.501.2677D} and \citetalias{2021MNRAS.501.4534J}.

We use optical spectroscopy from the Library of Libraries of Massive-Star High-Resolution Spectra project \citep[LiLiMaRlin; ][]{2019hsax.conf..420M}, which has an extensive collection of archival optical spectra. We include the following datasets:       

 \textit{NoMaDs}: We use the same spectroscopic observations (R=30,000) from the NoMaDs Project \citep{2012ASPC..465..484M} that were used in the period analysis of \citetalias{2012Wade.NGC.1624.2}. These observations were obtained between 2011 and 2012 with the HRS spectrograph on the Hobby Eberly Telescope (9.2 m) at the McDonald Observatory.  For our analysis, we use observations with the B (wavelength range of $\sim$380 to  470 nm), V ($\sim$530 to 630 nm), and R ($\sim$640 to 730 nm) configurations. 


 \textit{HERMES}: We use 2 high-resolution spectra gathered with the High Efficiency and Resolution Mercator Echelle Spectrograph (HERMES; R=85,000) on the Mercator telescope (1.2 m) at the Observatorio del Roque de Los Muchachos. These observations were obtained in October 2016 and 2017 and provide a wavelength coverage of $\sim$370 nm to 900 nm. 

 \textit{FIES}: We use observations obtained with the FIbre-fed \'{E}chelle Spectrograph (FIES) aboard the Nordic Optical Telescope at the ORM. These observations are made available in the OB star spectroscopic database from the IACOB project \citep{2015hsa8.conf..576S, 2020sea..confE.187S}. We use 4 observations, one obtained with the FIES high resolution mode (R=67,000) and 3 with the lower resolution mode (R=25,000). These observations were taken between November 2017 and October 2019 and provide a wavelength range of $\sim$370 nm to 900 nm.

 \textit{CAF\'{E}-BEANS}: We use four high-resolution (R=65,000) observations from the CAF\'{E}-BEANS (CAF\'{E}-Binary Evolution Andalusian Northern Survey) survey  \citep{2015hsa8.conf..524N}. This survey obtained observations from the Calar Alto Fiber-fed \'{E}chelle  spectrograph (CAF\'{E}) on the 2.2m telescope at the Calar Alto Observatory between October 2012 and January 2014. The observations provide a wavelength coverage from $\sim$360 nm to 920 nm. 

 \textit{DAO:} We use spectra obtained with the SITe-2 Cassegrain Spectrograph mounted on the 1.8m telescope at the Dominion Astrophysical Observatory (DAO), taken between November 2019 and February 2022. These R=15,700 observations are centered on the rest wavelength of the \hei\ line, with coverage from 574 nm to 600 nm.

For the ESPaDOnS and DAO observations, we co-add the spectra that were acquired in a single observing night and use their mean HJDs.  Overall, we have a multi-epoch collection of observations for time series analyses.

\section{Evidence for the need of a new period}
\label{sec:need_new_period}

\begin{figure*}
\centering
	\includegraphics[width=0.8\textwidth]{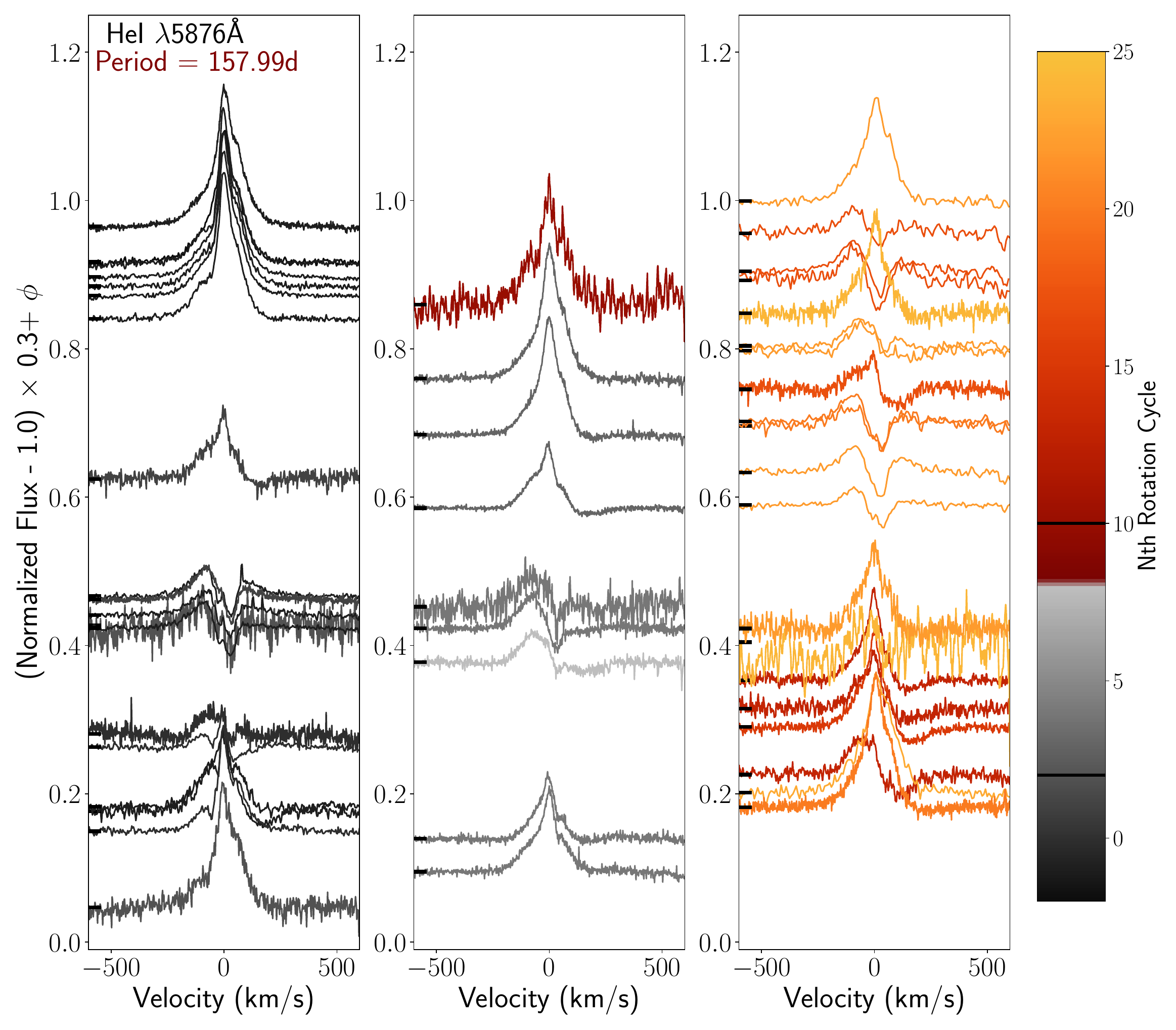}
    \caption{The \hei\ line profile variability. These normalized line profiles are plotted as a function of velocity in the stellar rest frame, scaled by 0.3, and shifted vertically according to their rotational phases using the ephemeris of \citetalias{2012Wade.NGC.1624.2} (the black dashes on the left are aligned with their continuum). The profiles are colored by their rotation number ($N_{\textrm{rot}}$). The profiles are split among the left, center, and right panels by $N_{\textrm{rot}} < 2 $, $ 2 \le  N_{\textrm{rot}} < 10  $, and $N_{\textrm{rot}} \ge 10$, respectively (also indicated by the black lines on the color bar). For display purposes, the MERCATOR observations were smoothed with a 7 point boxcar averaging.} 
    \label{fig:HE1_FIG}
\end{figure*}

In Figure \ref{fig:HE1_FIG}, we show the phase evolution of the \hei\ profiles when folded with a 157.99 d period (the 3 panels correspond to separate 3 epochs). We calculate the phase using the ephemeris reported by \citetalias{2012Wade.NGC.1624.2}, 

\begin{equation}
 \mathrm{HJD}  = 2455967.0 + 157.99 \times E \textrm{.}
\label{eq:phase}
\end{equation}

In Equation \ref{eq:phase}, $\mathrm{HJD}$ is the Heliocentric Julian Date of the observation, the `time-zero' Julian Date ($\mathrm{HJD}_0$) is $2455967.0 \pm 10.0$ (the $\mathrm{HJD}_0$ uncertainty represents the phase-shift added to align a high state with $\phi =0$), the rotational period (denoted by the variable $P$) is $157.99 \pm 0.94 \textrm{d}$, and  $E$ is the sum of the rotation number and phase. We calculate the phase uncertainty using conventional error propagation while using the uncertainties in $P$. As seen in the left panel of the figure (the first few rotation cycles around the magnetic field detection), \hei\ profiles transition from emission (at high state) to a complex morphology with partial emission (at low state) and then back to emission. At later epochs (right panel), the profiles no longer vary smoothly with their adopted phases. At these later rotation cycles, the complex line morphologies share the same phases as the emission profiles at earlier epochs. This demonstrates the need to revise the rotational period of \dashtwo.

\section{Methods} 
\label{sec:methods}


Our objective is to determine the rotational period of \dashtwo\ by examining the periodic variability of magnetospheric spectral lines. Following the method of \citetalias{2012Wade.NGC.1624.2}, we use the line profiles of \hei, \heii, \ha, and \hb\ from the observations discussed in Section \ref{sec:observations} and separately characterize their periodic variability with Lomb-Scargle (LS; \citealp{1976Ap&SS..39..447L}; \citealp{1982ApJ...263..835S}) periodograms. We use the LS periodogram because it is well-suited for time series with uneven and sparse sampling. 
\hei\ is the only line of the four that transitions into a complex morphology (hence why we chose it for visual displays); the others are consistently broad emission lines with variable intensities.

We apply the Lomb-Scargle method to time series of equivalent width (EW).  We will first introduce the general preparations of the data for this method. We will then discuss the relevant details and considerations in the LS periodogram analyses.

The ESPaDOnS and NARVAL observations were normalized by \citet{macinnis2016}. For the ESPaDOnS and NARVAL observations, the \ha\ and \hb\ spectral lines lie in the overlapping regions of neighboring spectral orders. We use the upper order for \ha\ and the lower order for \hb. Prior to co-addition, we use a 3rd-order polynomial fit of the continuum of each observation to normalize them. The remaining observations from the LiLiMaRlin project \citep{2019hsax.conf..420M} were already normalized.

We use the uncertainty-weighted implementations of LS, but the ESPaDOnS and NARVAL observations are the only ones that have pre-calculated flux error bars. We therefore measure the standard deviation of the normalized continuum and use it as a uniform uncertainty across all points on the line profile. To identify the continuum, we clip everything outside  5 $\sigma$ ($\sigma$ is the standard deviation of the full spectrum) and then apply a 2 $\sigma$ clip for the LiLiMaRlin observations. We apply the same initial sigma-clip to the DAO observations and use 3 $\sigma$ for the second clipping.


For each spectral line, we convert wavelengths to Doppler velocities and shift to NGC 1624-2's rest frame using its 34 km/s radial velocity (\citealp{2017MNRAS.465.2432G}). For each spectral line, we apply an additional local normalization using a linear fit on points selected from the continuum on each side of the spectral lines.

Using the locally normalized line profiles, we calculate the equivalent widths in velocity units. We use an integration range with symmetric bounds about $v=0$. We set the integration limits at 280, 300, 500, and 250 $km s^{-1}$ for \heii\, \hei\, \ha\, and \hb\, respectively. For the error bar calculation, we treat the integration as a right-handed Riemann approximation and apply conventional uncertainty propagation.

We analyze the variation of each equivalent width time series using LS periodograms \citep{1976Ap&SS..39..447L} computed with \texttt{astropy} \citep{astropy:2013, astropy:2018, astropy:2022}. Our grid of test periods, ranging from 10 d to 350 d, is uniformly spaced with $\Delta P =0.01$ d. Following standard considerations \citep{2018ApJS..236...16V}, this grid does not go below twice the shortest time interval between 2 consecutive EW measurements (effectively an approximation of the Nyquist Frequency when inverted), nor does it go above the total temporal length of each time series. We find that the minimum testable periods are 1.99 d, 1.87 d, 9.12 d, and 1.99 d for \heii\, \hei\, \ha\ and \hb\, respectively.

The \texttt{astropy} LS algorithm allows us to manage the complexity of the model, which is defined as:
\begin{equation}
y(t; f) = \theta_{0} + \sum_{n=1}^{N}[\theta_{2n -1} \sin(2 \pi n ft) + \theta_{2n} \cos(2 \pi n ft) ]. 
\label{eq:ls_eq}
\end{equation}
In this model, the free parameters are: the  midline $\theta_0$ (we use the mean EWs of each time series) and the amplitudes $\theta_{2n-1}$ and $\theta_{2n}$. The 1-term model ($N=1$) would be a simple sinusoidal curve, and using high-order Fourier series would mean progressively adding sinusoids with harmonic frequencies.

\begin{figure}[h]
  \centering
  \gridline{\fig{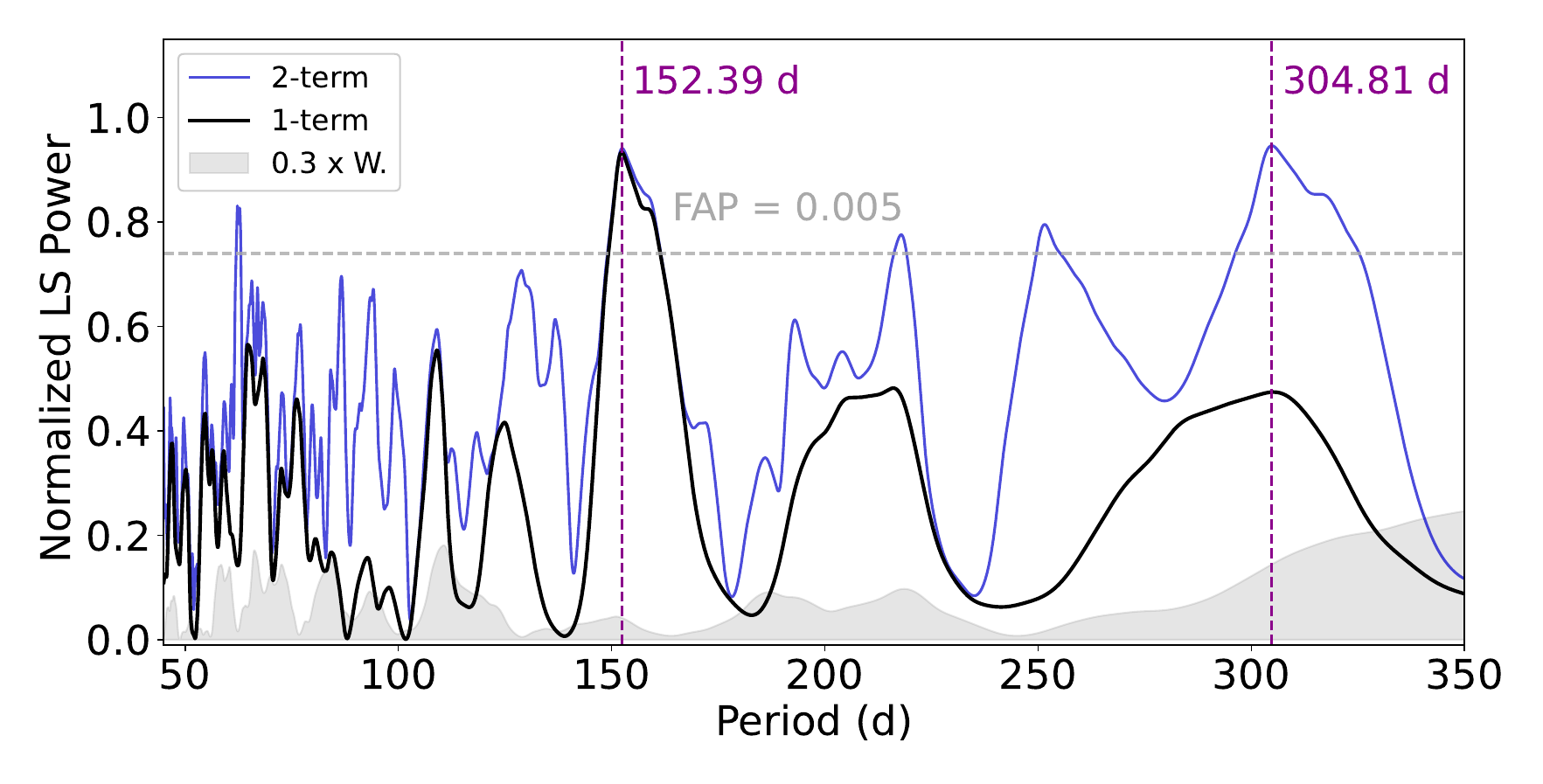}{\columnwidth}{(a)}}
  \gridline{\fig{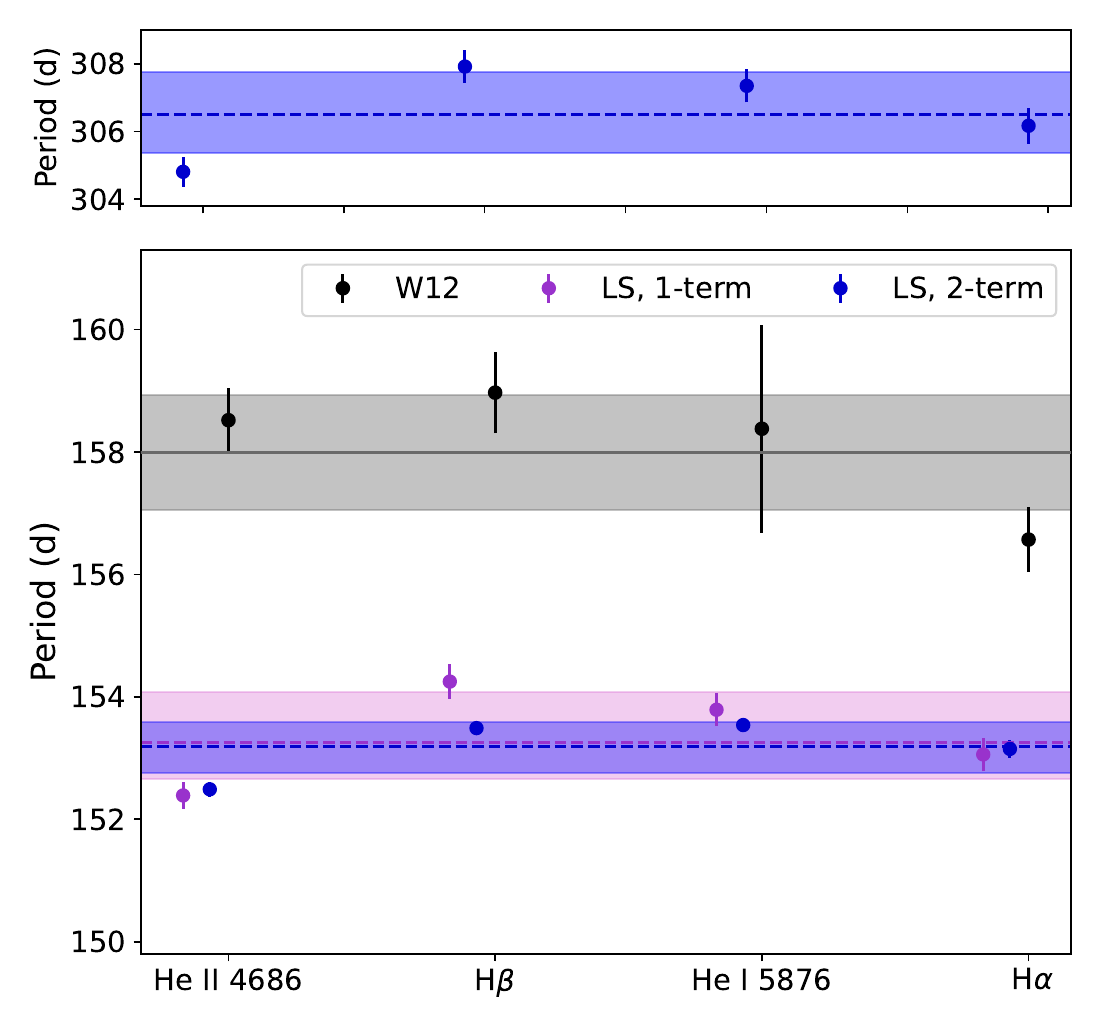}{\columnwidth}{(b)}}
    
\caption{a) Periodograms from the \heii\ time series. The dashed gray line indicates the power above which the False Alarm Probability is below 0.005 for the 1-term LS only. We also show the window spectrum scaled by a factor of 0.3 in gray. We display the periodograms as a function of period to be consistent with Figure 5 of \citetalias{2012Wade.NGC.1624.2}. b) Significant periods identified in the periodogram of the four spectral lines. The  black points are measurements from the \citetalias{2012Wade.NGC.1624.2}. The dark gray line and shaded area represent the original $157.99\pm0.94$ d period. The purple and blue points are our 1-term and 2-term Lomb-Scargle (LS) derived periods, respectively. The dashed lines and shaded areas show the standard mean of the periods and their associated standard deviations.\label{fig:pdgram_plus_composit}}
\end{figure}

We compute both 1-term and 2-term LS periodograms. We provide an example of these periodograms for the \heii\ time series in Figure \ref{fig:pdgram_plus_composit}(a). In the periodograms, we identify the peak with the highest power and verify that the peak is above the level at which the False Alarm Probability (FAP) is at most 0.005 (indicated by the gray dashed line)\footnote{The FAP quantifies the probability that a peak would occur at or above a certain threshold, given the null hypothesis that the time series is not periodic. Peaks with high FAP are likely spurious.}. We calculate the FAP's via the bootstrap implementation in \texttt{astropy}'s Lomb-Scargle function, but only for the 1-term LS (neither bootstrap nor analytical FAP estimates are available for the 2-term LS method). The bootstrap method randomly samples the EWs while keeping the original timestamps and computes a distribution of LS periodograms. \citet{2018AJ....156...82C} states that "the bootstrap produces the most robust estimate of the FAP because it makes few assumptions about the form of the periodogram distribution and fully accounts for survey window effects."  
 
As shown in Figure \ref{fig:pdgram_plus_composit}(a), we identify $P = 152.39$ d as a significant period from the 1-term periodogram and 152.49 d from the 2-term periodogram. 
The 2-term periodogram also has large power at 304.81 d. Indeed, the 1-term and 2-term periodograms have significantly different values of power near $\sim 300$ d, which is a harmonic of the $\sim 150$ d period. This difference is primarily due to the model and the way magnetospheric spectral lines can vary. The 1-term periodogram evaluates the least-squares fit for a single-wave sinusoidal model, while the 2-term fit can produce a double-wave model for the same period.
The 1-term and 2-term periodograms both assign high power to the single-wave period around $\sim150$ d. For double-wave periods, the 1-term LS attempts to fit a single wave that spans $\sim$300 days, while the 2-term LS fits combinations of $\sim$300 d and $\sim$150 d waves (which better reproduce double high state cycles). Thus, the 2-term LS results in higher power around double wave periods than the 1-term. 

For the uncertainty of the LS-derived periods, we use the analytic approximation provided by \citet{1986ApJ...302..757H}. All of the single-wave and double-wave periods with high LS power are listed in Table \ref{tab:table_all_periods}.

\section{Results}
\label{sec:results}

In this section, we discuss i) the periods that produce single-wave variations (one emission maximum per cycle) and ii) the periods that produce double-wave variations (two emission maxima per cycle and approximately twice as long as the single-wave). 
Subsequently, we assess the manner in which the EW time series phases according to these periods and investigate the periods' consistency with the ORM.    

\subsection{Periods yielding `single-wave' variations }
\label{sec:results_sw_periods}

Figure \ref{fig:pdgram_plus_composit}(b) visually compares the periods reported in Table \ref{tab:table_all_periods} to those from \citetalias{2012Wade.NGC.1624.2}. Our objective is to determine a period that is overall compatible with the variations of all the spectral lines.

\begin{table}
 \caption{The significant periods (in d) that we identified in each LS periodogram. We compute the means and standard deviations of single-wave periods and the double-wave. We also include the lines wavelengths from NIST's Atomic Spectra Database \citep{2000AIPC..543..299W}.}
 
\begin{tabular}{lcc}
\hline
Spectral Line & \multicolumn{2}{c}{LS Period} \\   
  &  1-term & 2-term  \\
\hline
\multicolumn{3}{l}{\textbf{Single-wave}}\\
He \textsc{ii} $\lambda$4685.568  & 152.39 $\pm$ 0.22 & 152.49 $\pm$ 0.12  \\
He \textsc{i} $\lambda$5875.621 & 153.79 $\pm$ 0.27 & 153.54 $\pm$ 0.12  \\
H$\alpha$ $\lambda$6562.819 & 153.06 $\pm$ 0.28 & 153.15 $\pm$ 0.15  \\
H$\beta$  $\lambda$4861.350  & 154.25 $\pm$  0.28 & 153.49 $\pm$ 0.11  \\

Mean $\pm$ Std Dev & 153.37 $\pm$ 0.71 & 153.17 $\pm$ 0.42 \\
\hline
\multicolumn{3}{l}{\textbf{Double-wave}}\\
He \textsc{ii} $\lambda$4685.568  & - & 304.81 $\pm$ 0.44 \\
He \textsc{i} $\lambda$5875.621 & -  & 307.35 $\pm$ 0.49  \\
H$\alpha$ $\lambda$6562.819 & - & 306.17 $\pm$ 0.54  \\
H$\beta$  $\lambda$4861.350  & - & 307.92 $\pm$ 0.48 \\

Mean $\pm$ Std Dev & - & 306.56 $\pm$ 1.19 \\
\hline
\end{tabular}

 \label{tab:table_all_periods}
\end{table}

For the single-wave periods, the LS periodograms of each spectral line have significant peaks between 152.39 and 154.25 days, which are a few days shorter than the 157.99 d period of \citetalias{2012Wade.NGC.1624.2}. When comparing the values for each individual spectral line, it is apparent that all of our periods disagree with the values of \citetalias{2012Wade.NGC.1624.2} for a given spectral line. 

For our 1-term LS periods, \heii\ and \hb\ agree at the 4 $\sigma$ level, while other pairings agree within 2 to 3 $\sigma$. For the 2-term LS periods, \heii\ has a 5 $\sigma$ agreement with \hb\ and \hei\, while other pairings agree within 1 to 3 $\sigma$. For each spectral line, 1-term and 2-term counterparts agree at the 1 to 2 $\sigma$ level.

\begin{figure*}
\centering

	\includegraphics[width=0.95\textwidth]{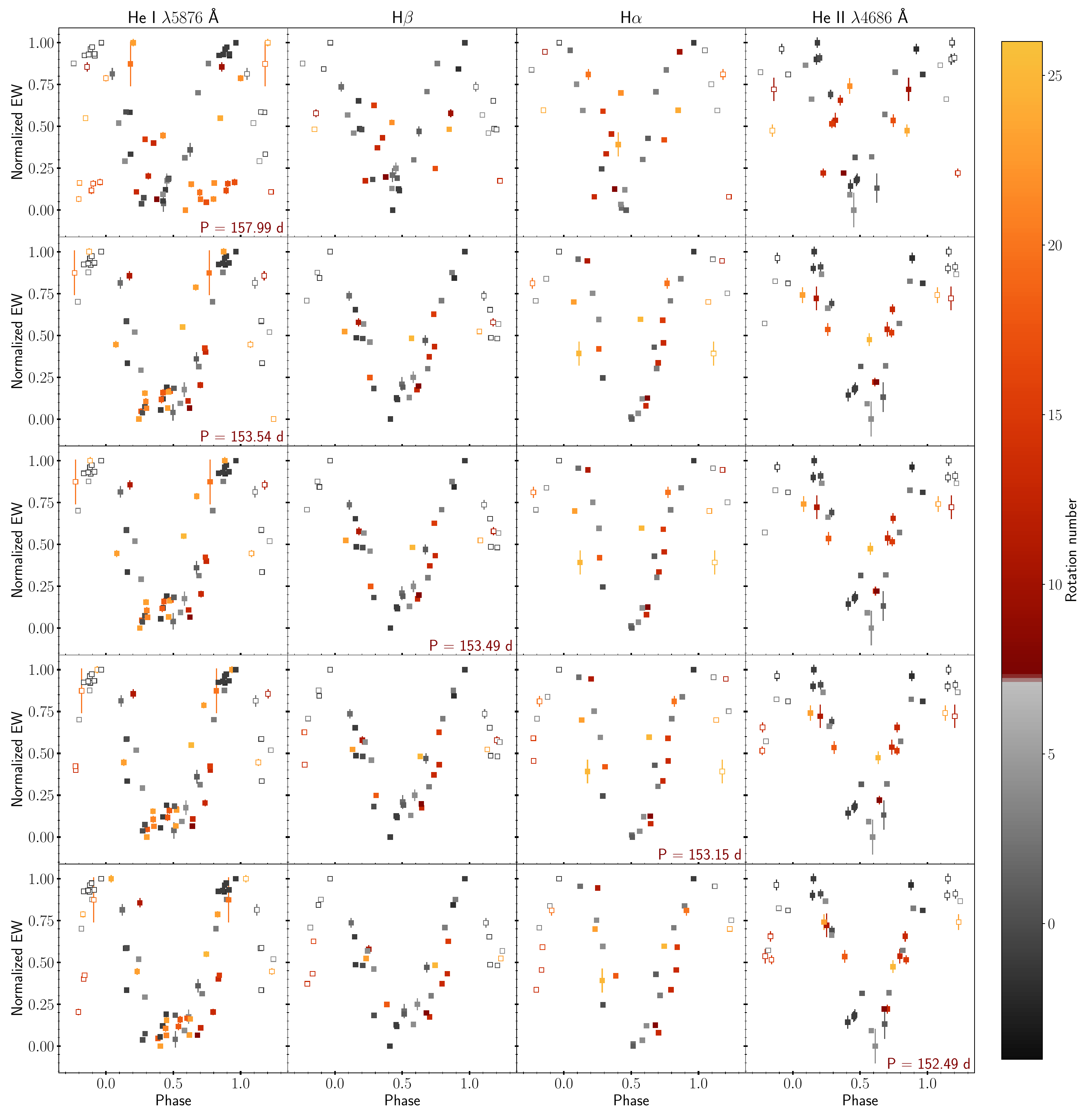}
    \caption{The variability of each EW time series using different periods. In the upper panels, the time series are phase-folded according to the \citetalias{2012Wade.NGC.1624.2} rotational period. In descending order of row, the four time series are phased by 153.54 d, 153.49 d, 153.15 d, and 152.49 d, which are all 2-term LS periods. The time series are vertically flipped and  normalized so that 1 represents the most negative EWs (i.e. strongest emission/high state) and 0 indicates the opposite (ie. weakest emission/ low state; recall that the low state He \textsc{i} profiles are complex). The measurements are colored according to their rotation numbers. To visually highlight the variation, we add unfilled duplicates of the first and last quarters of the time series.
    \label{fig:ew_w12_vs_all}}
\end{figure*}


We assess how effectively the periods phase the EW time series \footnote{We use the same HJD$_0$ as \citetalias{2012Wade.NGC.1624.2} in Equation \ref{eq:phase}.} and determine which ones produce the most coherent folded patterns. Figure \ref{fig:ew_w12_vs_all} shows the collective phasing of each time series by a given period. We compare the 157.99 d period (upper row of panels) with the 2-term LS periods.

When we apply the \citetalias{2012Wade.NGC.1624.2} period, all the phase-folded time series are disordered: the EWs of the newer observations (red and orange points) are significantly out of phase with the earliest ones (gray-scaled points). This is most apparent for the \hei\ and \heii\ EWs. 

Applying the LS periods to their respective time series (i.e., the main diagonal of the bottom 4 rows) results in more coherent variability. For these periods, the measurements conform to a single-wave sinusoidal pattern, with maximum emission near $\phi= 0.0$ and minimum emission near $\phi= 0.5$. 
The early epochs (black to gray) align more closely with the later ones (red to yellow).

We also examine how each of these four LS periods phases the EW of the other spectral lines (e.g., how well does the period derived from \hb\ phase the EWs of \ha, \hei, and \heii). 
Comparing the largest period ($P =$ 153.54 d) with the other periods reveals that the EW measurements converge toward more coherent sinusoidal patterns as the period decreases. The 153.15 d and 152.49 d periods yield the most compact groupings between early- and late-epoch observations. 

Additionally, there is notable scatter that none of the adopted periods could effectively remove. 
The measurements from rotations 10 to 15 (red) appear more coherent with $P=153.17$ d, whereas measurements from rotations 20 to 25 (orange) appear more coherent with $P=152.49$ d.

This behavior may be the consequence of cycle-to-cycle fluctuations in the magnetospheric emission. \citet{2013MNRAS.428.2723U} produced 3D MHD simulations of a stellar magnetosphere similar to the magnetic O-type star $\theta^1$ Ori C. They quantified the cycle-to-cycle scatter in observed and synthetic H$\alpha$ EWs, and found that both had a standard deviation of $\sim$0.25 \AA\ about a harmonic fit. With EWs variations from -1 \AA\ to 2 \AA\ over rotational phase, this standard deviation corresponds $\sim$0.1 in normalized units.

We adopt the same approach to quantify the cycle-to-cycle scatter in our EWs, over our optimal range of periods. We fit a harmonic functions (same form as Eq. \ref{eq:ls_eq}) with N = 2, 3, 4, find the coefficients with least-squares fitting, and compute the standard deviations of the residuals. Considering all of the lines together, the residual scatter is $\sim$0.15. 
This value is much larger than the mean uncertainty of the normalized EWs ($\sim$ 0.02 level), but is comparable to \citet{2013MNRAS.428.2723U}. Increasing the number of harmonics does not significantly reduce the scatter, and the standard deviation remains nearly constant across our adopted period uncertainty.

Overall, a period 4 to 6 days shorter than that of \citetalias{2012Wade.NGC.1624.2} produces more coherent, sinusoidal, modulated variations in the equivalent width measurements. Accordingly, for the remainder of our analysis, we adopt the mean and standard deviation of the 2-term LS periods, $P = 153.17 \textrm{d} \pm 0.42 \textrm{d}$ (see the 153.15 d period shown in Figure \ref{fig:ew_w12_vs_all} for a comparable reference). 

\begin{figure*}[!ht]
\centering
 \gridline{\fig{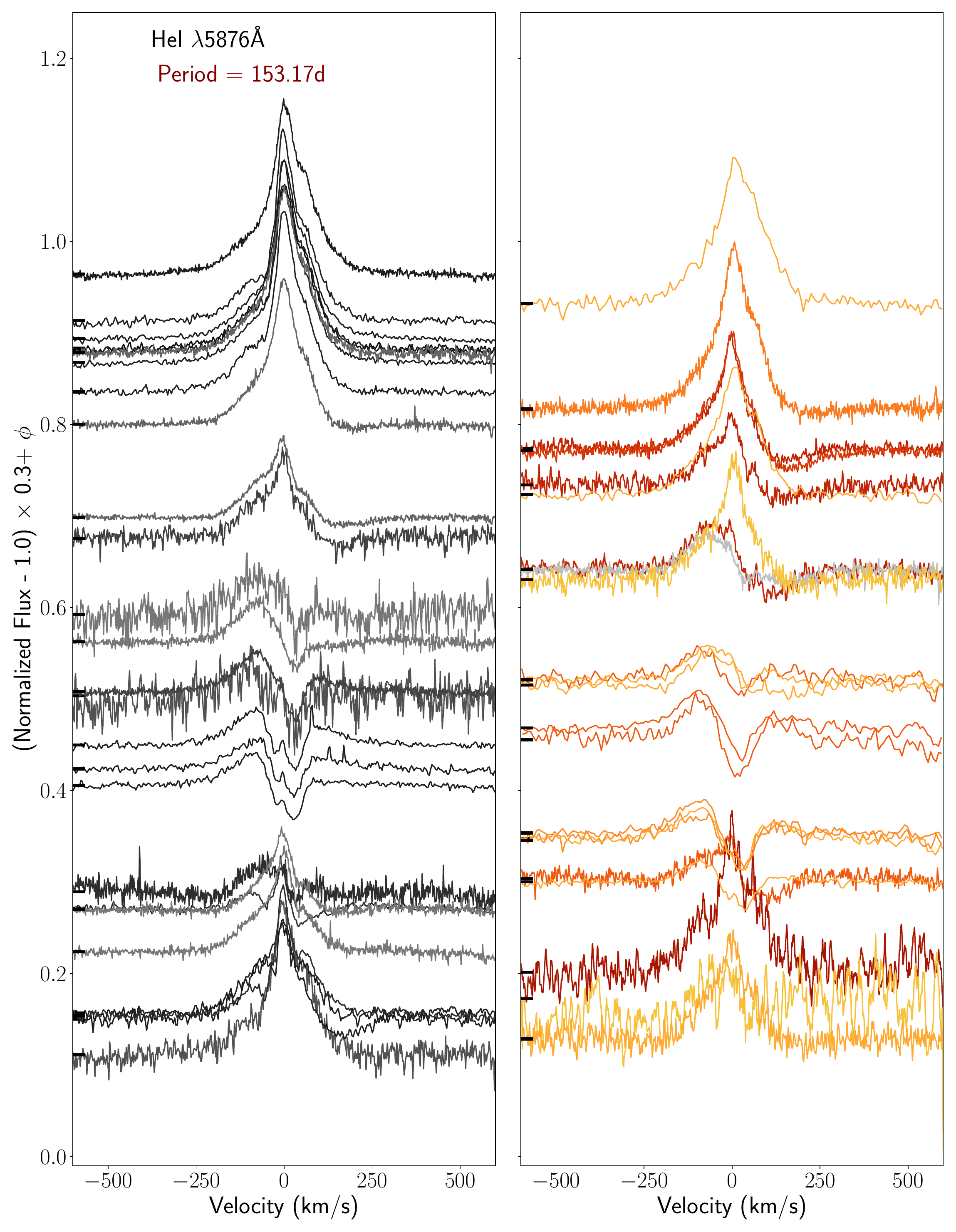}{0.46\textwidth}{(a)}
  \fig{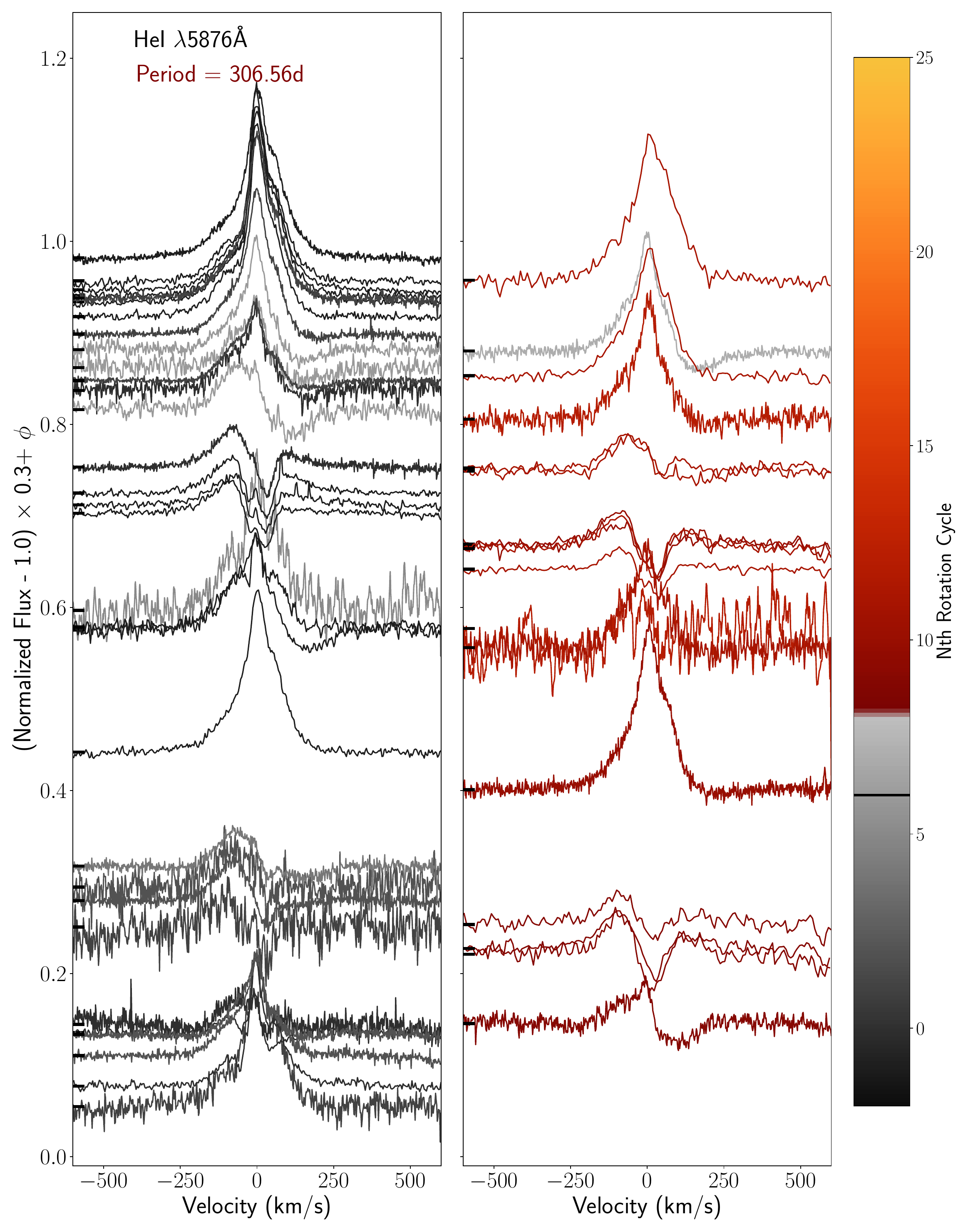}{0.465\textwidth}{(b)}}
 
    \caption{a) Same as Figure \ref{fig:HE1_FIG}, but the panels use $P =$ 153.17 d to phase the profiles. b) Same as Figure \ref{fig:HE1_FIG}, but the panels use $P =$ 306.56 d. Both figures share the same color, and the profiles are divided into the panels by $N_{\textrm{rot}} < 6 $.}
    \label{fig:he_prof_153_306}
\end{figure*}

In Figure \ref{fig:he_prof_153_306}(a), we directly examine the variations of the He\textsc{i} profiles as a function of phase using the adopted 153.17 d period. This approach allows us to identify any significant cycle-to-cycle discrepancies that are not apparent in the equivalent widths. The leftmost panel, which contains early-epoch observations, resembles that of Figure \ref{fig:HE1_FIG}. This is expected, as this period is only slightly shorter than 157.99 d, and not enough time has elapsed (since HJD$_0$) for significant phase drifting. 

The second panel shows late-epoch observations. We find that the phasing of the profiles is consistent with early rotation cycles. Most notably, the He\textsc{i} profiles with complex morphologies appear to be confined to a phase interval of 0.3 to 0.7 for each epoch. Some  discrepancies persist, such as near $\phi \sim$ 0.7. We find that these discrepancies do not outweigh the overall improvements from using this period. The uncertainty in phase ($\delta\phi$) increases as the time displacement relative to HJD$_0$ does. For the observations in the second panel of Figure \ref{fig:he_prof_153_306}(a), $\delta\phi$ $\sim$ 0.03 for some of the earliest observations (at $N_{\textrm{rot}}= 10$), and $\delta\phi$ $\sim$ 0.07 for the latest observation (at $N_{\textrm{rot}}= 25$). Overall, from our set of single-wave periods, 153.17 d offers an optimal characterization of the variability of magnetospheric lines for NGC 1624-2.

\subsubsection{The variation of \texorpdfstring{$\langle B_\textrm{z} \rangle$\ }\  with a "single-wave" period}
\label{sec:results_sw_mag}

\begin{figure*}
\centering
    \gridline{\fig{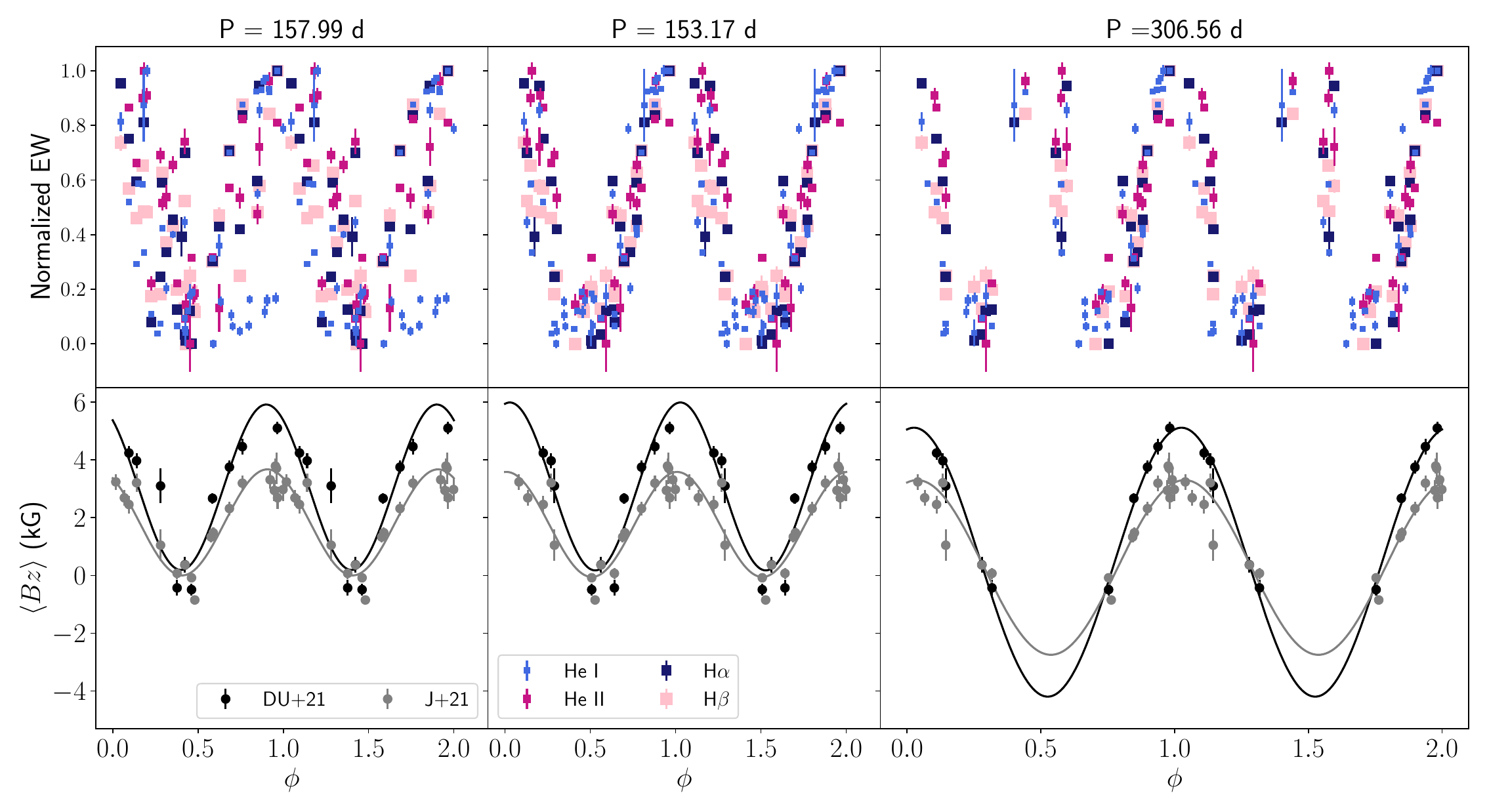}{0.79\textwidth}{(a)}
          \fig{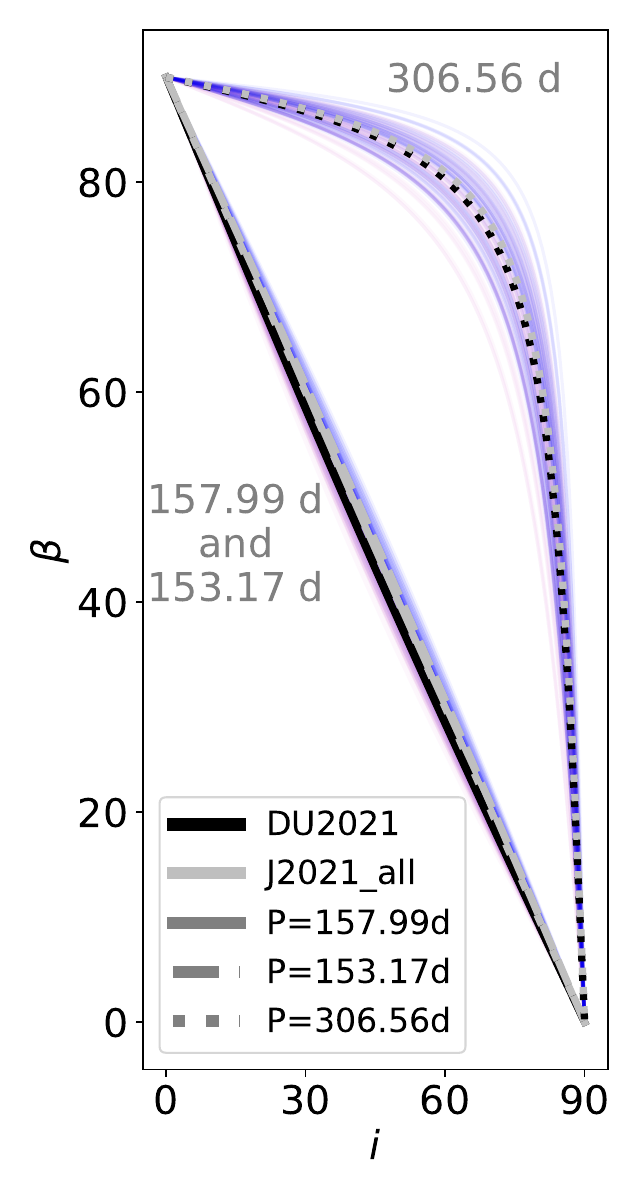}{0.21\textwidth}{(b)}}

    \caption{a)The variability of the normalized EW time series (He\textsc{i}, He\textsc{ii}, H$\alpha$, and H$\beta$; top panels) and the $\langle B_\textrm{z} \rangle$ measurements from \citetalias{2021MNRAS.501.2677D} and \citetalias{2021MNRAS.501.4534J} (bottom panels) using the 157.99 d, 153.17 d, and 306.56 d periods. The solid curves are 1st-order sinusoidal fits to the $\langle B_\textrm{z} \rangle$ measurements using each period. The data are replicated over two rotation cycles for visual enhancement. b) Relationship between $i$ and $\beta$ that is implied by the min-to-max ratio of the best fitting sinusoid curves to the $\langle B_\textrm{z} \rangle$ measurements. The three types of dashed curves show the $i$-$\beta$ curves derived for the \citetalias{2021MNRAS.501.2677D} (black) and \citetalias{2021MNRAS.501.4534J} (gray) datasets using each period. The colored curves are the $i-\beta$ curves resulting from samples drawn from the MCMC fit posterior distribution; the pink and purple curves are from fits to \citetalias{2021MNRAS.501.2677D} and \citetalias{2021MNRAS.501.4534J}, respectively; the set of lower curves is from using the single-wave periods, and the set of higher curves is from the double-wave period.
    }
    \label{fig:ORM_plus_ibeta}
\end{figure*}

Next, we confirm whether the 153.17 d period is compatible with the variability of the magnetic field measurements. We apply this revised ephemeris to the independent sets of $\langle B_\textrm{z} \rangle$ measurements reported by \citetalias{2021MNRAS.501.2677D} and \citetalias{2021MNRAS.501.4534J}.These measurements were made with the same ESPaDOnS and NARVAL observations that we included in our spectroscopic analysis \footnote{We note that \citetalias{2021MNRAS.501.4534J} has a larger set of measurements than \citetalias{2021MNRAS.501.2677D} as they did not co-add the ESPaDOnS observations taken during the same night. We opted to use the measurements from \citetalias{2021MNRAS.501.4534J} that included all applicable lines in the Least-Squares Deconvolution profiles}. In Figure \ref{fig:ORM_plus_ibeta}(a), we compare the EW variations of all the  spectral lines (top panels) and the $\langle B_\textrm{z} \rangle$ measurements (bottom panels). We include sinusoidal fits to the $\langle B_\textrm{z} \rangle$ data; the fit parameters, the amplitude, mean (vertical offset), and phase offset, are listed in Table \ref{tab:cos_fit}.

 The left panels of Figure \ref{fig:ORM_plus_ibeta}(a) show the measurements phased with the 157.99 day period, and the middle panels show them phased with the $P =$ 153.17 d period. The right panel shows phasing with $P =$ 306.56 d, which we will discuss in Section \ref{sec:results_dw_mag}.

 Using the 153.17 d period instead of 157.99 d yields only minor changes to the phases and sinusoidal fits to the $\langle B_\textrm{z} \rangle$ data. The magnetic measurements do not undergo large phase shifts from changing periods, as they are mostly from early-epoch observations. The $\langle B_\textrm{z} \rangle$ fits maxima remain near 6 kG (4 kG for \citetalias{2021MNRAS.501.4534J}) while the minima remain near 0 kG or very weakly negative. Overall, with the 153.17 day period, we find no contradiction between the variations of spectral lines and magnetic field measurements with respect to the ORM.

In summary, the single-wave periods that we obtain are all shorter than the 157.99 d period. When applied to equivalent widths, these shorter periods reduce the phase offsets that were observed in recent observations while using the 157.99 d period. We adopt the mean 2-term LS period of 153.17 d, which minimizes phase drifts of all spectral lines; although cycle-to-cycle variability persists. This shorter period also remains compatible with the magnetic datasets of \citetalias{2021MNRAS.501.2677D} and \citetalias{2021MNRAS.501.4534J}.

\subsection{Periods yielding "double-wave" variations  }
\label{sec:results_dw_periods}

We now consider the possibility of a period roughly twice that found in the previous section. As mentioned in Section \ref{sec:intro}, if the star has, e.g., a highly oblique dipolar field, then both poles can be significantly visible during a rotation cycle, and we would observe two high states over the course of one rotation.

The upper panel of Figure \ref{fig:pdgram_plus_composit}(b) presents a visual comparison of the significant double-wave periods. The periods derived from \heii\ and \hb\ agree at the 4$\sigma$ level, while remaining pairings agree within 2$\sigma$. Overall, these periods are $\sim$6 to $\sim$12 d shorter than twice the \citetalias{2012Wade.NGC.1624.2} period (315.98 $\pm$ 1.88 d). We adopt the mean and standard deviation of the double-wave periods, $P = 306.56\ \textrm{d} \pm 1.19\ \textrm{d}$.

We do not replicate Figure \ref{fig:ew_w12_vs_all} for the double-wave periods, as the phasing of the EW time series is similar for these periods. Using either the shortest(304.81 d) or longest (307.92 d) double-wave period instead of 306.56 d yields a maximum phase shift of 0.06 on the last observation, which represents the largest possible phase offset. We therefore use the upper right panel of Figure \ref{fig:ORM_plus_ibeta}(a) to show the impact of using a 306.56 d period on the EW variations. 
This 306.56 d period meets our expectations for double-wave variations of magnetospheric spectral lines. Two emission maxima (high states) occur within one cycle at $\phi \sim 0.0$ and $\phi \sim 0.5$. There are also two emission minima (low states) at $\phi \sim 0.25$ and $\phi \sim 0.75$.  
Although the two high-states appear to have similar amplitudes, the sampling of the mid-cycle high state is not sufficient to definitively determine this. 
We measure the cycle-to-cycle scatter in the same way as for the single-wave periods, and we get very similar results: there is constant residual scatter at the 0.15 level, in the range of period around our nominal double-wave period.

As was done in Section \ref{sec:results_sw_periods}, we examine the variation of He\textsc{i} profiles when phase-folded with the 306.56 d period, as shown by Figure \ref{fig:he_prof_153_306}(b). In both panels, the complex He\textsc{i} profiles are grouped around $\phi$ $\sim$0.25 and $\phi \sim$ 0.75.
Strong emission profiles occur at the start of the cycle ($\phi \sim$ 0.0) as well as at $\phi \sim$ 0.5. 
The smooth transitions between strong emission and complex morphologies are consistent with the high states and low states observed in the EW variation in Figure \ref{fig:ORM_plus_ibeta}(a). 
Some phasing discrepancies remain, such as at $\phi$ $\sim$ 0.8. 
However, for the observation in the right panel of Figure \ref{fig:he_prof_153_306}(b), $\delta\phi$ $\sim$ 0.02 at $N_{\textrm{rot}}$ = 6 and $\delta\phi$ $\sim$ 0.05 for the most recent observation ($N_{\textrm{rot}}$ = 12). Overall, the variability of the spectral lines appear compatible with a 306.56 d double-wave period.

\subsubsection{The variation of \texorpdfstring{$\langle B_\textrm{z} \rangle$\ }\  with a "double-wave" period}
\label{sec:results_dw_mag}
We examine the co-variations of the EW and $\langle B_\textrm{z} \rangle$ measurements when phased with the 306.56 d period. 
In Figure \ref{fig:ORM_plus_ibeta}(a) (lower right panel), the $\langle B_\textrm{z} \rangle$ measurements from both studies are well fitted by the sinusoidal curves. 
The quality of the $\langle B_\textrm{z} \rangle$ fits improves significantly when adopting the longer period (see reduced $\chi^2$ values listed in Table \ref{tab:cos_fit}). 
We find that the variations of the $\langle B_\textrm{z} \rangle$ measurements (and their sinusoidal models) are consistent with the EW datasets, with extrema of $\langle B_\textrm{z} \rangle$ occurring at the same phase as the EW emission maxima, as expected. 
At the start of the cycle ($\phi\sim$0.0), the mean longitudinal field reaches its strongest positive value; it is $\langle B_\textrm{z} \rangle 
\ \sim$ 5.11 kG and 3.30 kG for the \citetalias{2021MNRAS.501.2677D} and \citetalias{2021MNRAS.501.4534J} datasets, respectively. 
The near-zero $\langle B_\textrm{z} \rangle$ measurements are instead distributed between two separate EW low states at $\phi \sim0.25$ and $\sim0.75$. 

The sinusoidal models also predict the magnetic South pole approaching the line of sight, yielding strongly negative minima (of $-4.19$ kG for the \citetalias{2021MNRAS.501.2677D} measurements and $-2.74$ kG for \citetalias{2021MNRAS.501.4534J}) that coincide with an EW high state at $\phi \sim 0.5$. 
Overall, the variability of the EW and $\langle B_\textrm{z} \rangle$ measurements is consistent with $P = 306.56$ d within the framework of the ORM and magnetospheric emission.

Importantly, no existing spectropolarimetric observations are within the phase interval during which the double-wave models predict strongly negative $\langle B_\textrm{z} \rangle$ measurements. Consequently, the existing longitudinal field dataset is unable to distinguish between the single-wave and double-wave solutions.

\section{Discussion and  Conclusions}
\label{sec:disc_conc}
We analyzed the variability of magnetospheric spectral lines of NGC 1624-2 in order to update the measurement of the rotation period. We used the LS method on EW measurements of spectral lines from both new and archival observations. 
Additionally, we assessed how well the identified periods phased the EWs, the line profiles, and the literature $\langle B_\textrm{z} \rangle$ measurements (from \citetalias{2021MNRAS.501.2677D} and \citetalias{2021MNRAS.501.4534J}).   
From the periodograms, we found periods that phase the EW time series into single-wave and double-wave variations. 
We adopted $153.17 \pm 0.42 \textrm{d}$ as the single-wave period and $306.52 \pm 1.19$ d as the double-wave period. 

Both periods are effective in reducing the late-epoch phase drifts that were present whenever the 157.99 d period \citepalias{2012Wade.NGC.1624.2} was used. 
The single-wave or double-wave periods produce coherent variations in the EW and line profile time series. 

We found that some scatter remained present in the phased equivalent width measurements, regardless of which period we used. 
This is not entirely surprising, as cycle-to-cycle fluctuations are observed in other magnetic O stars (HD 148937, Of?p; \citealp{2012MNRAS.419.2459W} and $\theta^1$ Ori C, O7f?p; \citealp{2008A&A...487..323S, 2019A&A...626A..20M}). 
The investigation by \citet{2013MNRAS.428.2723U} into the stochasticity of magnetospheric \ha\ emission from $\theta^1$ Ori C found that over-densities of in-falling material were the main sources of the fluctuations and contributed to the azimuthal fragmentation of breakout outflows above the Alfv\'{e}n radius.Our measurement of the scatter in our EWs is at the same level as that reported by \citet{2013MNRAS.428.2723U}, and they do not differ significantly between single and double-wave periods. Because the harmonic fits return similar residuals, this suggests that i) cycle-to-cycle variations are present, and ii) the variation of the EWs cannot distinguish between the single-wave and double-wave periods.

\subsection{Magnetic Geometry}
\label{sec:disc_mag_geo}

The variations of the $\langle B_\textrm{z} \rangle$ measurements reported by \citetalias{2021MNRAS.501.2677D} and \citetalias{2021MNRAS.501.4534J} are compatible with both rotational periods, thus confirming consistency with the ORM. 
However, the double-wave period reveals a sizable gap in spectropolarimetric coverage, which leaves us unable to definitively determine which period is correct. This gap is likely a dual-consequence of the visibility window from CFHT and the timing of observations. If the star actually completes its rotations in just under a year and the north pole was in view at discovery, the south pole would not have been visible from CFHT for a few years post-discovery.

The two possible rotational periods correspond to very different magnetic configurations.

The \textbf{single-wave period} corresponds to a configuration in which only the North magnetic pole comes into view during a stellar rotation. As shown in Figure \ref{fig:ORM_plus_ibeta}(a), our single-wave period does not significantly impact the phasing of the spectropolarimetric measurements, as they were all taken at an early epoch. Thus, our result for the single-wave variations unsurprisingly matches the configuration inferred by \citetalias{2012Wade.NGC.1624.2} (and revisited by  \citetalias{2021MNRAS.501.2677D} and \citetalias{2021MNRAS.501.4534J}). 

We provide Figure \ref{fig:ORM_plus_ibeta}(b) to show the constraints between $i$ and $\beta$ implied by the sinusoidal fits to the $\langle B_\textrm{z} \rangle$ data in Figure \ref{fig:ORM_plus_ibeta}(a), as described by \citet{1967ApJ...150..547P}. We derive the Preston $r$-value which is the ratio of $\langle B_\textrm{z} \rangle$ extrema ($r = B_{min} / B_{max}$). After calculating $r$, we evaluate $i$ with respect to $\beta$ using $\tan(i) = ((1-r)/(1+r)) \cot(\beta)$. Each curve represents a Markov Chain Monte Carlo (MCMC) sample in the posterior probability. Broadly, the fits have a Preston $r$-value close to zero, and the magnetic configuration is such that $i + \beta$ is roughly 90 degrees.

The \textbf{double-wave period} corresponds to a configuration in which both poles are visible. As shown in Figure \ref{fig:ORM_plus_ibeta}(a), adopting this period significantly alters the phases of spectropolarimetric measurements. Our model reproduces the measurements well and reaffirms the observation of the northern magnetic hemisphere. However, due to the gap in coverage, we also interpret the EW curves.
For double-wave variation, the high state peaks of the EWs can differ in amplitude depending on the relative tilt of the two magnetic poles when each is closest to the line of sight. When using the double-wave period for \dashtwo, we do not observe any clear asymmetries between the EW high states. That said, the mid-cycle high state is not fully covered, even with our new spectroscopic observations. At present, the high state appears to have an amplitude at least comparable to the one coinciding with the North magnetic pole. This suggests that the $\langle B_\textrm{z} \rangle$\ variation is more symmetric about zero, implying a magnetic field that may be more tilted than originally inferred.

Our fit of the $\langle B_\textrm{z} \rangle$\ data phased with the double-wave period supports a large obliquity or inclination angle. As seen in Figure \ref{fig:ORM_plus_ibeta}(a), the $\langle B_\textrm{z} \rangle$ fit predicts extrema of similar strengths, with a Preston $r$-value close to -1. 
Figure \ref{fig:ORM_plus_ibeta}(b) shows the resulting constraints on $i$ and $\beta$ for this period. Broadly, either $\beta$ has to be close to 90 degrees  (for any non-zero values of $i$) or $i$ has to be close to 90 degrees (for any non-zero values of $\beta$). Within these constraints, smaller values of $i$ or $\beta$ require a stronger dipolar field to reproduce the $\langle B_\textrm{z} \rangle$ extrema. 

For a 306 d period, an estimated stellar radius of $10 \textrm{R}_{\odot}$ yields an equatorial rotational velocity of $\sim$1.5 km/s, consistent with the inferred $v\sin i$ < 3 km/s. However, this offers no meaningful constraint on the inclination angle.

At present, we are unable to distinguish between the single- and double-wave periods with only EWs. Given the available data and the level of cycle-to-cycle variability, it is unlikely that spectroscopy alone can resolve this ambiguity, especially if the magnetic geometry results in emission maxima of similar amplitude. 
The most straightforward approach to discern between the single- and double-wave periods is to obtain at least one Stokes $V$ observation within the phase interval where the double-wave period predicts that the (yet unobserved) South pole would come into view.

\subsection{Revisiting Phasing in Previous Studies}
\label{sec:past_studies}

We now address the implications of the two rotation periods on previous magnetospheric studies of NGC 1624-2. 
We specifically revisit the UV observations reported by \citetalias{2021MNRAS.501.2677D}, and the X-ray analysis of \citet{2015MNRAS.453.3288P}.  

\subsubsection{UV Observations}
\label{sec:UV_analysis}

\citet{2019MNRAS.483.2814D, 2021MNRAS.501.2677D} observed the UV resonance spectral lines formed in the winds of O stars. As wind lines are usually very sensitive to the density and kinematics of the stellar outflows, the break in spherical symmetry caused by the magnetosphere is expected to have a significant impact on the shape of the line profiles \citep{2014A&A...568A..59S, 2021MNRAS.506.5373E}. They found that the variations of the C\textsc{iv} $\lambda\lambda$ 1548, 1550 \AA\ and S\textsc{iv} $\lambda\lambda$ 1393, 1402 \AA\ doublets were quite extreme, and that their shapes were not compatible with spherically symmetric winds. In light of our revised periods, we revisit their findings in which the variation of the line profiles as a function of phase was not what one would expect from the magnetic geometry assumed at the time.

\begin{figure*}
\centering
  \includegraphics[width=\textwidth]{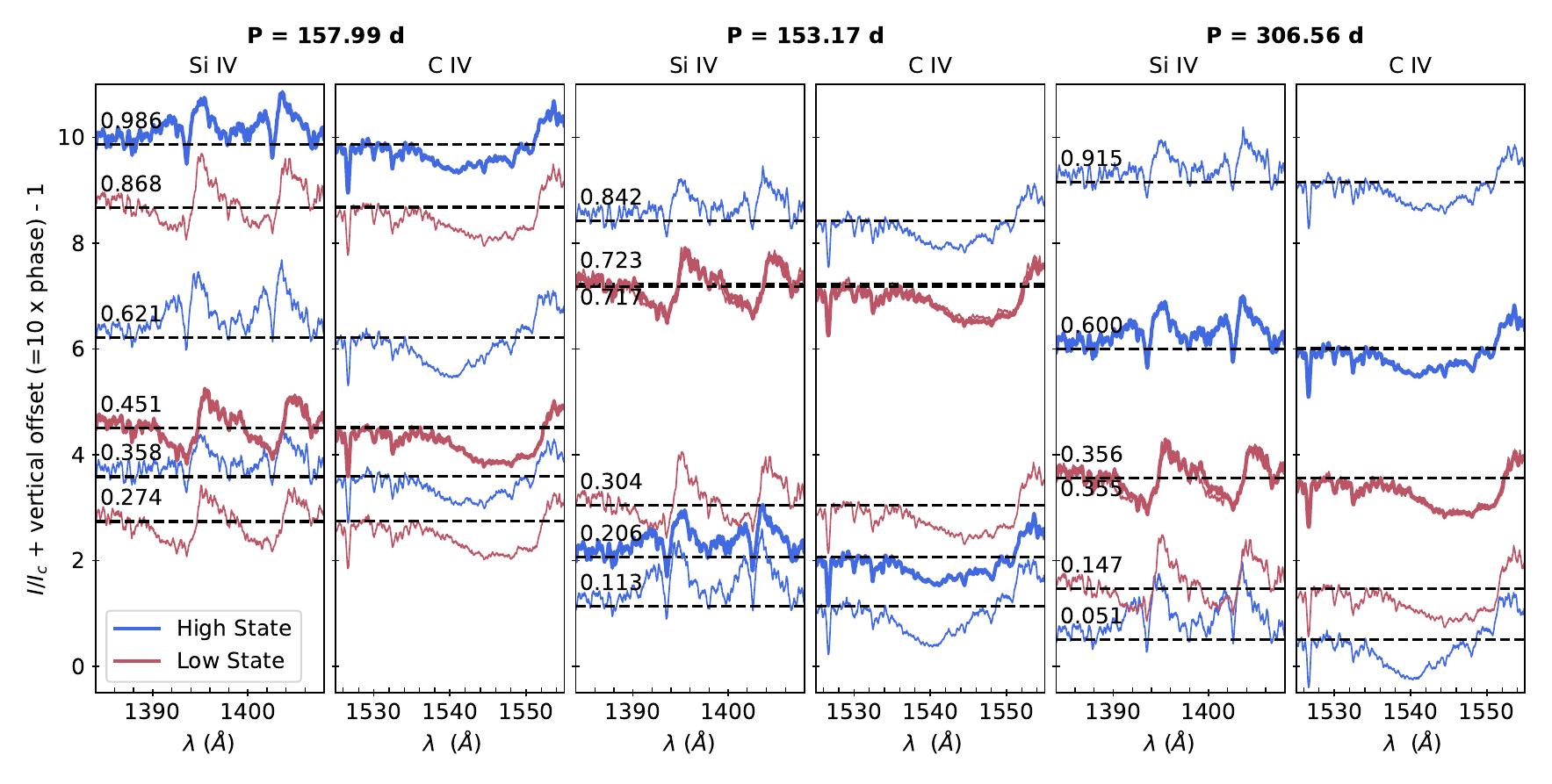}
    \caption{Profiles of the C\textsc{iv} and Si\textsc{iv} wind-sensitive resonance lines for NGC 1624-2 that were studied by \citetalias{2021MNRAS.501.2677D}. In the same manner as Figure \ref{fig:HE1_FIG}, the profiles are arranged vertically according to phase (scaled up by a factor of 10 as in \citetalias{2021MNRAS.501.2677D}). The observations with thicker lines are the earlier-epoch high state (blue) and low state (red) that were reported in \citet{2019MNRAS.483.2814D}. The others are follow-up observations with the color assigned by \citetalias{2021MNRAS.501.2677D} according to which of the two reference observations they resembled the most. In the middle and right panels, we phase the observations according to our single-wave and double-wave ephemerides.}
    \label{fig:UV_plots}
\end{figure*}

In Figure \ref{fig:UV_plots} (left panel), we reproduce Figure 1 of \citetalias{2021MNRAS.501.2677D}. In this figure, the initial observations by \citet{2019MNRAS.483.2814D} (bold) served as a reference to categorize the follow-up spectra by their morphology. Blue and red indicate observations that have similar line morphologies as the reference high state and low state observations, respectively.
\citetalias{2021MNRAS.501.2677D} focused on C\textsc{iv} and Si\textsc{iv} doublets, given their value as magnetospheric probes (we refer to \citetalias{2021MNRAS.501.2677D} and \citealt{2015MNRAS.452.2641N} for details). 

With a single-wave period, \citetalias{2021MNRAS.501.2677D} expected the line profile morphology to transition from a high state near phase 0 to a low state at phase 0.5 and back to a high state at phase 1.0. However, as seen in the leftmost panels of the figure, they found that the 157.99 d period does not phase the UV observations in a continuous single-wave variation.

In the middle panel of Figure \ref{fig:UV_plots}, we apply our single-wave ephemeris to phase the UV data. There is now a consistent, smooth transition from high state-like to low state-like morphologies.
The shift in phasing of the UV observations is large because they were obtained at late epochs with respect to HJD$_0$ (rotations 14 and 16) and have accumulated a larger phase difference than, e.g. the archival spectropolarimetric observations. In other words, these observations were not obtained at their intended rotational phases. 

The right panel of Figure \ref{fig:UV_plots} shows the UV observations phased with the 306.56 d period. As expected, we see a double-wave variation of the spectral lines. 
With this period, two of the observations would correspond to the North high state, and one to the South high state. 
Two observations would correspond to nearly the same rotational phase ($\sim0.35$), and the agreement between the two observations is very good.

\subsubsection{X-ray Observations}
\label{sec:xray_analysis}

 \citet{2015MNRAS.453.3288P} conducted a follow-up X-ray analysis to \citetalias{2012Wade.NGC.1624.2} with two Chandra observations targeting the high state and low state, as predicted with the \citetalias{2012Wade.NGC.1624.2} ephemeris. They found variable emission that could only be modeled with a strong source of soft X-rays that is heavily attenuated at low energies. This finding is consistent with a model in which X-rays are generated by magnetically confined wind shocks \citep{1997A&A...323..121B, 2014MNRAS.441.3600U} and are partially reabsorbed by the extensive, cooled magnetospheric material.

We confirm that these observations captured their intended magnetospheric states when adopting either of our derived rotation periods. These observations were acquired within a maximum of 4 rotations from the zero-date of the ephemeris, so the accumulated phase offsets are minimal.
The low state observation’s (HJD 2456508.20) phase changes from 0.43 to 0.53 with our 153.17 d period and to 0.77 with the 306.56 d period (which is the low state following the visibility of the negative magnetic pole). 
The high state observation’s (HJD 2456592.84) phase changes from 0.96 to 0.09 with the 153.17 d period and to 0.04 with the 306.56 d period (which would be a North high state observation).

\subsection{Conclusion}
\label{sec:conclusion}

We have identified two periods that can simultaneously phase the magnetospheric and magnetic measurements of \dashtwo\ and reopened questions regarding the star's magnetic field configuration. 

Our analysis of magnetospheric spectral line time series has led us to a single-wave period of 153.17 d, which is only a few days shorter than the current rotation period on record. Using the currently accepted 157.99 d period would result today in an accumulated phase error of $\sim$ 0.16. 
Our shorter period still represents the originally inferred geometry that brings only one pole into view over the course of the rotation. This new single-wave period can reliably be used to schedule observations for studies that do not rely on the polarity of the high state. 

We also find that 306.56 days, which implies a geometry that allows both poles to be visible, is an equally suitable period. A spectropolarimetric observation obtained at the expected South pole high state would straightforwardly distinguish between a single-wave or a double-wave magnetic configuration. 

\section*{Statement of Data Availability}
The observations are available as follows:
{\it HST} COS data are archived under MAST: \dataset[10.17909/d8mf-2v58]{http://dx.doi.org/10.17909/d8mf-2v58}, ESPaDOnS and DAO via the Canadian Astronomical Data Centre, Narval via\dataset[Polarbase]{https://www.polarbase.ovgso.fr/}\citep{2014PASP..126..469P}, and the remaining data in their respective project repositories (or requested from the survey first authors).The optical observations and analysis are available in ZENODO via \url{https://doi.org/10.5281/zenodo.17953046}. 

\section*{Acknowledgements}

SS acknowledges support for this work from the Delaware Space Grant College and Fellowship Program (NASA
Grant 80NSSC20M0045).

This material is based upon work supported by the National Science Foundation under Grant No. AST-2108455 (SS, VP, JM).
MEO gratefully acknowledges support for this work from the National Science Foundation under Grant No. AST-2107871.

The authors would like to extend our thanks to the anonymous referees who provided numerous helpful comments that improved this paper.  

This work was made possible by observations obtained with the Canada–France–Hawaii Telescope (CFHT), which is operated by the National Research Council (NRC) of Canada, the Institut National des Sciences de l’Univers of the Centre National de la Recherche Scientifique (CNRS) of France, and the University of Hawaii. This work was made possible by observations obtained with the Narval spectropolarimeter at the Telescope Bernard Lyot, Observatoire du Pic du Midi, CNRS/INSU, and Université de Toulouse, France. This work was made possible by observations obtained with the HERMES spectrograph, which is supported by the Research Foundation - Flanders (FWO), Belgium, the Research Council of KU Leuven, Belgium, the Fonds National de la Recherche Scientifique (F.R.S.-FNRS), Belgium, the Royal Observatory of Belgium, the Observatoire de Genève, Switzerland, and the Thüringer Landessternwarte Tautenburg, Germany. This work was made possible by observations collected in the IACOB spectroscopic database maintained by the Nordic Optical Telescope Scientific Association using the Nordic Optical Telescope. This work was made possible by observations collected with the Mercator Telescope  on the island of La Palma at the Spanish Observatorio del Roque de los Muchachos of the Instituto de Astrofísica de Canarias. This work was made possible by observations obtained by the Dominion Astrophysical Observatory, which is part of the Herzberg Astronomy and Astrophysics Research Centre, under the National Research Council of Canada. This work was made possible by observations obtained by the Hobby-Eberly Telescope (HET) at the Texas McDonald Observatory, a joint project of the University of Texas at Austin, the Pennsylvania State University, Ludwig-Maximilians-Universität München, and Georg-August-Universität Göttingen. This work was made possible by observations collected at the Calar Alto Observatory. This work was made possible by observations made with the NASA/ESA Hubble Space Telescope obtained from the Space Telescope Science Institute, which is operated by the Association of Universities for Research in Astronomy, Inc., under NASA contract NAS 5–26555. These observations are associated with program HST-GO-15066.001-A. This research has made use of data obtained from the Chandra Data Archive and the Chandra Source Catalog, both provided by the Chandra X-ray Center (CXC).


\appendix

\section{The log of observations} 
\label{sec:append_obs_log}

\restartappendixnumbering
Table \ref{tab:tab_obs_log} provides a log of all observations used. We estimate the SNR's using a portion of the continuum on the red side of each spectral line. We include the equivalent widths that comprise each time series. We exclude a MERCATOR (2022-08-31)  observation from the calculation of any spectral time series due to high noise. We omit two CAFE BEANS ( 2012-12-28 and 2013-02-26) observations from the \heii\ time series and one (2020-11-02) from both the \heii\ and \hb\ time series due to significant noise. Due to evidence of saturation, we omit all NoMaDs observations from the \ha\ time series.

\begin{longtable}{llrl|rrrr|rrrrrrrr}
 \caption{\label{tab:tab_obs_log} The log of observations used in this work. We list the date, instrument, and exposure time, SNR's and EW measurements from each observation. We use shortened labels indicate the instruments: ESPaDOnS (ESP), NARVAL (NAR), NoMaDs(NOM), HERMES (MER), FIES (IAC), CAF\'{E}-BEANS (CAF). Observations acquired by the same instrument on the same night (or within one day of each other) are coadded, and we adopt the mean HJD for the spectral line time series. Some sets of ESPaDOnS observations are the exception: we coadd the observations acquired between 2012-02-01 and 2012-02-09 and those acquired between 2013-08-27 and 2013-08-29}  \\

\hline

Date & Obs. &  HJD &  Exp. (s) & \multicolumn{4}{c}{Signal-to-Noise Ratios} &  \multicolumn{8}{c}{Equivalent Width Measurements (km/s)}  \\
 & & - 2450000.0 & & He \textsc{i}\ & He \textsc{ii}\ &  H$\alpha$ &   H$\beta$ &  He \textsc{i} &  $\sigma$ & He \textsc{ii} &  $\sigma$ &   H$\alpha$ & $\sigma$ &  H$\beta$ &  $\sigma$ \\
\hline
                2011-08-22 & NOM (BGRV) & 5795.963 &  650 &    101.0 &     117.0 &        77.0 &       60.0 &   -84.6 &         0.6 &    -63.0 &          1.3 &        - &            - &   -241.21 &          0.46 \\
                2011-10-02 & NOM (BGRV) & 5836.866 &  900 &     53.0 &     102.0 &        18.0 &       59.0 &   -54.6 &         0.9 &    -60.5 &          1.3 &        - &            - &   -220.73 &          0.54 \\
                2011-10-03 & NOM (BGRV) & 5837.860 &  900 &     74.0 &     147.0 &        91.0 &       57.0 &   -32.4 &         0.8 &    -64.5 &          1.2 &        - &            - &   -202.71 &          0.40 \\
                2011-11-10 &   NOM (RV) & 5875.953 &  900 &    116.0 &         - &        87.0 &          - &    -7.6 &         0.5 &        - &            - &        - &            - &         - &             - \\
                2011-11-11 &   NOM (BG) & 5876.716 &   1800 &        - &      95.0 &           - &       60.0 &       - &           - &    -30.3 &          1.6 &        - &            - &   -150.37 &          0.48 \\
                2011-11-13 &   NOM (RV) & 5878.708 &  900 &    107.0 &         - &        81.0 &          - &   -13.4 &         0.6 &        - &            - &        - &            - &         - &             - \\
                2011-11-17 &   NOM (RV) & 5882.707 &  900 &    118.0 &         - &        68.0 &          - &   -19.6 &         0.5 &        - &            - &        - &            - &         - &             - \\
                2011-11-18 &   NOM (BG) & 5883.697 &   1800 &        - &     114.0 &           - &       63.0 &       - &           - &    -31.7 &          1.3 &        - &            - &   -164.01 &          0.44 \\
                2011-11-19 &   NOM (BG) & 5884.691 &   1800 &        - &     113.0 &           - &       64.0 &       - &           - &    -32.0 &          1.4 &        - &            - &   -162.91 &          0.44 \\
                2012-01-15 &   NOM (VR) & 5941.783 &  900 &    122.0 &         - &        84.0 &          - &   -84.9 &         0.6 &        - &            - &        - &            - &         - &             - \\
                2012-01-20 &   NOM (VR) & 5946.755 &    900 &    146.0 &         - &        67.0 &          - &   -85.4 &         0.5 &        - &            - &        - &            - &         - &             - \\
                2012-01-22 &   NOM (VR) & 5948.749 &    900 &    110.0 &         - &        50.0 &          - &   -88.1 &         0.6 &        - &            - &        - &            - &         - &             - \\
                2012-01-24 &   NOM (VR) & 5950.756 &    900 &    153.0 &         - &        93.0 &          - &   -89.2 &         0.5 &        - &            - &        - &            - &         - &             - \\
                2012-01-27 &   NOM (VR) & 5953.743 &      900 &    100.0 &         - &        47.0 &          - &   -85.7 &         0.6 &        - &            - &        - &            - &         - &             - \\
                2012-02-01 &        ESP & 5958.715 &     2400 &    138.0 &     141.0 &        93.0 &       59.0 &   -91.6 &         0.2 &    -56.9 &          0.3 &  -1236.3 &          0.4 &   -258.13 &          0.33 \\
                2012-02-02 &        ESP & 5959.716 &     2400 &        - &         - &           - &          - &       - &           - &        - &            - &        - &            - &         - &             - \\
                2012-02-03 &        ESP & 5960.713 &     2400 &        - &         - &           - &          - &       - &           - &        - &            - &        - &            - &         - &             - \\
                2012-02-04 &        ESP & 5961.713 &     2400 &        - &         - &           - &          - &       - &           - &        - &            - &        - &            - &         - &             - \\
                2012-02-09 &        ESP & 5966.720 &     2400 &        - &         - &           - &          - &       - &           - &        - &            - &        - &            - &         - &             - \\
                2012-03-04 &   NOM (VR) & 5990.621 &      900 &    100.0 &         - &        42.0 &          - &   -54.9 &         0.6 &        - &            - &        - &            - &         - &             - \\
                2012-03-12 &   NOM (BG) & 5998.600 &   1800 &        - &     153.0 &           - &       62.0 &       - &           - &    -60.9 &          1.1 &        - &            - &   -202.25 &          0.41 \\
                2012-03-22 &   NOM (VR) & 6008.596 &      900 &    130.0 &         - &        63.0 &          - &    -6.1 &         0.6 &        - &            - &        - &            - &         - &             - \\
                2013-03-24 &        NAR & 6011.332 &     4800 &     36.0 &      30.0 &        30.0 &       31.0 &    -9.3 &         0.7 &    -52.1 &          1.1 &   -888.8 &          1.1 &   -170.05 &          0.97 \\
                2012-09-27 &        ESP & 6197.944 &     5400 &    142.0 &     114.0 &        82.0 &       60.0 &   -19.1 &         0.2 &    -37.2 &          0.3 &   -775.4 &          0.3 &   -170.95 &          0.30 \\
                2012-09-27 &        ESP & 6198.009 &     5400 &        - &         - &           - &          - &       - &           - &        - &            - &        - &            - &         - &             - \\
       2012-10-23 &      CAF & 6223.633 &   3600 &     36.0 &      22.0 &        41.0 &       19.0 &   -34.7 &         3.6 &    -29.9 &          3.5 &   -973.1 &          4.8 &   -201.01 &          3.36 \\
       2012-12-28 &      CAF & 6290.400 &   3600 &     32.0 &      19.0 &        31.0 &       12.0 &   -75.0 &         3.1 &        - &            - &  -1215.4 &          4.1 &   -229.71 &          3.14 \\
       2013-02-26 &      CAF & 6350.464 &   3600 &     20.0 &       8.0 &        18.0 &        8.0 &    -6.3 &         4.4 &        - &            - &   -781.2 &          5.8 &   -172.89 &          4.50 \\
                2013-08-27 &        ESP & 6532.119 &     5160 &    167.0 &     115.0 &        94.0 &       53.0 &   -30.5 &         0.2 &    -37.3 &          0.2 &   -914.8 &          0.3 &   -182.75 &          0.26 \\
                2013-08-29 &        ESP & 6534.056 &     5160 &        - &         - &           - &          - &       - &           - &        - &            - &        - &            - &         - &             - \\
                2013-08-29 &        ESP & 6534.118 &     5160 &        - &         - &           - &          - &       - &           - &        - &            - &        - &            - &         - &             - \\
                2013-09-13 &        ESP & 6549.033 &     5160 &    140.0 &     106.0 &        71.0 &       50.0 &   -65.0 &         0.2 &    -47.4 &          0.3 &  -1101.2 &          0.3 &   -226.62 &          0.31 \\
                2013-09-13 &        ESP & 6549.096 &     5160 &        - &         - &           - &          - &       - &           - &        - &            - &        - &            - &         - &             - \\
                2013-09-25 &        ESP & 6561.024 &     5160 &    115.0 &      93.0 &        66.0 &       52.0 &   -80.6 &         0.3 &    -57.4 &          0.3 &  -1161.6 &          0.4 &   -244.74 &          0.37 \\
                2013-09-25 &        ESP & 6561.086 &     5160 &        - &         - &           - &          - &       - &           - &        - &            - &        - &            - &         - &             - \\
                2013-11-17 &        ESP & 6613.909 &     5160 &    124.0 &     107.0 &        65.0 &       56.0 &   -48.9 &         0.3 &    -59.1 &          0.3 &  -1121.8 &          0.4 &   -211.57 &          0.37 \\
                2013-11-17 &        ESP & 6614.041 &     5160 &        - &         - &           - &          - &       - &           - &        - &            - &        - &            - &         - &             - \\
                2013-11-24 &        ESP & 6621.036 &     5160 &    121.0 &      87.0 &        73.0 &       53.0 &   -28.7 &         0.3 &    -51.0 &          0.3 &  -1049.8 &          0.4 &   -199.91 &          0.37 \\
                2013-11-24 &        ESP & 6621.098 &     5160 &        - &         - &           - &          - &       - &           - &        - &            - &        - &            - &         - &             - \\
                2014-01-08 &        ESP & 6665.822 &     5160 &     91.0 &      87.0 &        71.0 &       51.0 &   -11.1 &         0.3 &    -28.3 &          0.3 &   -791.4 &          0.4 &   -164.27 &          0.40 \\
                2014-01-08 &        ESP & 6665.884 &     5160 &        - &         - &           - &          - &       - &           - &        - &            - &        - &            - &         - &             - \\
       2014-01-12 &      CAF & 6670.413 &   3600 &     23.0 &       8.0 &        23.0 &       10.0 &   -18.3 &         3.9 &    -24.7 &          4.1 &   -831.2 &          5.0 &   -177.26 &          3.80 \\
                2015-09-24 &        ESP & 7290.036 &     5160 &     91.0 &      84.0 &        46.0 &       48.0 &    -8.5 &         0.3 &    -33.5 &          0.4 &   -833.0 &          0.4 &   -171.64 &          0.44 \\
                2015-09-25 &        ESP & 7291.054 &     5160 &        - &         - &           - &          - &       - &           - &        - &            - &        - &            - &         - &             - \\
2016-10-21 &        MER & 7682.744 &   1800 &     14.0 &       8.0 &        14.0 &        6.0 &   -78.8 &         2.5 &    -53.4 &          2.8 &  -1211.0 &          3.1 &   -212.69 &          2.65 \\
2017-10-30 &        MER & 8056.556 &   3600 &     39.0 &      25.0 &        27.0 &       18.0 &   -12.4 &         0.9 &    -33.5 &          1.1 &   -812.0 &          1.2 &   -169.17 &          0.98 \\
       2017-11-13 &        IAC & 8070.611 &   3600 &     43.0 &      21.0 &        47.0 &       24.0 &   -20.8 &         1.8 &    -46.0 &          1.8 &   -930.3 &          2.3 &   -190.43 &          1.66 \\
       2017-11-19 &        IAC & 8076.660 &   5400 &     61.0 &      36.0 &        49.0 &       36.0 &   -38.3 &         1.2 &    -50.8 &          1.2 &   -985.0 &          1.6 &   -196.94 &          1.17 \\
       2018-09-21 &        IAC & 8382.717 &   3600 &     74.0 &      39.0 &        55.0 &       35.0 &   -40.4 &         1.0 &    -45.2 &          1.1 &  -1047.6 &          1.4 &   -217.81 &          1.01 \\
       2019-10-14 &        IAC & 8770.613 &   2400 &     47.0 &      25.0 &        37.0 &       26.0 &    -6.8 &         1.5 &    -45.9 &          1.5 &   -968.9 &          2.0 &   -177.16 &          1.46 \\
                2019-11-06 &        dao & 8793.798 &     1800 &     50.0 &         - &           - &          - &   -13.1 &         2.0 &        - &            - &        - &            - &         - &             - \\
                2019-11-06 &        dao & 8793.819 &     1800 &        - &         - &           - &          - &       - &           - &        - &            - &        - &            - &         - &             - \\
                2019-11-06 &        dao & 8793.841 &     1800 &        - &         - &           - &          - &       - &           - &        - &            - &        - &            - &         - &             - \\
                2019-11-06 &        dao & 8793.862 &     1800 &        - &         - &           - &          - &       - &           - &        - &            - &        - &            - &         - &             - \\
                2019-11-08 &        dao & 8795.760 &     1800 &    115.0 &         - &           - &          - &   -16.8 &         1.9 &        - &            - &        - &            - &         - &             - \\
                2019-11-08 &        dao & 8795.782 &     1800 &        - &         - &           - &          - &       - &           - &        - &            - &        - &            - &         - &             - \\
                2019-11-08 &        dao & 8795.803 &     1800 &        - &         - &           - &          - &       - &           - &        - &            - &        - &            - &         - &             - \\
                2019-11-08 &        dao & 8795.831 &     1800 &        - &         - &           - &          - &       - &           - &        - &            - &        - &            - &         - &             - \\
                2019-11-16 &        dao & 8803.778 &     1800 &     63.0 &         - &           - &          - &   -17.5 &         1.8 &        - &            - &        - &            - &         - &             - \\
                2019-11-16 &        dao & 8803.799 &     1800 &        - &         - &           - &          - &       - &           - &        - &            - &        - &            - &         - &             - \\
                2019-11-16 &        dao & 8803.821 &     1800 &        - &         - &           - &          - &       - &           - &        - &            - &        - &            - &         - &             - \\
       2020-11-02 &      CAF & 9155.513 &   1800 &        - &         - &           - &          - &       - &           - &        - &            - &        - &            - &         - &             - \\
                2021-01-22 &        dao & 9236.758 &     1800 &     67.0 &         - &           - &          - &   -12.1 &         1.9 &        - &            - &        - &            - &         - &             - \\
                2021-01-22 &        dao & 9236.789 &     1800 &        - &         - &           - &          - &       - &           - &        - &            - &        - &            - &         - &             - \\
                2021-01-22 &        dao & 9236.810 &     1800 &        - &         - &           - &          - &       - &           - &        - &            - &        - &            - &         - &             - \\
                2021-01-22 &        dao & 9236.831 &     1800 &        - &         - &           - &          - &       - &           - &        - &            - &        - &            - &         - &             - \\
                2021-01-23 &        dao & 9237.677 &     1800 &    106.0 &         - &           - &          - &    -8.4 &         1.3 &        - &            - &        - &            - &         - &             - \\
                2021-01-23 &        dao & 9237.698 &     1800 &        - &         - &           - &          - &       - &           - &        - &            - &        - &            - &         - &             - \\
                2021-01-23 &        dao & 9237.719 &     1800 &        - &         - &           - &          - &       - &           - &        - &            - &        - &            - &         - &             - \\
                2021-01-23 &        dao & 9237.740 &     1800 &        - &         - &           - &          - &       - &           - &        - &            - &        - &            - &         - &             - \\
                2021-01-23 &        dao & 9237.771 &     1800 &        - &         - &           - &          - &       - &           - &        - &            - &        - &            - &         - &             - \\
                2021-01-23 &        dao & 9237.792 &     1800 &        - &         - &           - &          - &       - &           - &        - &            - &        - &            - &         - &             - \\
   2021-10-22 &        IAC & 9509.608 &   1037 &     30.0 &      23.0 &        26.0 &       22.0 &   -42.3 &         1.9 &    -54.2 &          1.9 &  -1097.9 &          2.5 &   -206.78 &          1.79 \\
                2021-11-17 &        dao & 9535.886 &     1800 &    112.0 &         - &           - &          - &    -2.7 &         1.4 &        - &            - &        - &            - &         - &             - \\
                2021-11-17 &        dao & 9535.908 &     1800 &        - &         - &           - &          - &       - &           - &        - &            - &        - &            - &         - &             - \\
                2021-11-17 &        dao & 9535.929 &     1800 &        - &         - &           - &          - &       - &           - &        - &            - &        - &            - &         - &             - \\
                2021-11-24 &        dao & 9542.838 &     1800 &     99.0 &         - &           - &          - &   -16.5 &         1.6 &        - &            - &        - &            - &         - &             - \\
                2021-11-24 &        dao & 9542.859 &     1800 &        - &         - &           - &          - &       - &           - &        - &            - &        - &            - &         - &             - \\
                2021-12-20 &        dao & 9568.771 &     1800 &     61.0 &         - &           - &          - &    -8.6 &         1.5 &        - &            - &        - &            - &         - &             - \\
                2021-12-20 &        dao & 9568.792 &     1800 &        - &         - &           - &          - &       - &           - &        - &            - &        - &            - &         - &             - \\
                2021-12-20 &        dao & 9568.813 &     1800 &        - &         - &           - &          - &       - &           - &        - &            - &        - &            - &         - &             - \\
                2021-12-20 &        dao & 9568.834 &     1800 &        - &         - &           - &          - &       - &           - &        - &            - &        - &            - &         - &             - \\
                2021-12-21 &        dao & 9569.765 &     1800 &    105.0 &         - &           - &          - &   -17.1 &         1.4 &        - &            - &        - &            - &         - &             - \\
                2021-12-21 &        dao & 9569.786 &     1800 &        - &         - &           - &          - &       - &           - &        - &            - &        - &            - &         - &             - \\
                2021-12-21 &        dao & 9569.807 &     1800 &        - &         - &           - &          - &       - &           - &        - &            - &        - &            - &         - &             - \\
                2021-12-21 &        dao & 9569.828 &     1800 &        - &         - &           - &          - &       - &           - &        - &            - &        - &            - &         - &             - \\
                2022-01-21 &        dao & 9600.683 &     1800 &    126.0 &         - &           - &          - &   -72.7 &         1.7 &        - &            - &        - &            - &         - &             - \\
                2022-01-21 &        dao & 9600.704 &     1800 &        - &         - &           - &          - &       - &           - &        - &            - &        - &            - &         - &             - \\
                2022-01-21 &        dao & 9600.725 &     1800 &        - &         - &           - &          - &       - &           - &        - &            - &        - &            - &         - &             - \\
                2022-02-22 &        dao & 9632.629 &     1800 &     59.0 &         - &           - &          - &   -91.6 &         2.0 &        - &            - &        - &            - &         - &             - \\
                2022-02-22 &        dao & 9632.650 &     1800 &        - &         - &           - &          - &       - &           - &        - &            - &        - &            - &         - &             - \\
                2022-02-22 &        dao & 9632.671 &     1800 &        - &         - &           - &          - &       - &           - &        - &            - &        - &            - &         - &             - \\
                2022-02-22 &        dao & 9632.691 &     1800 &        - &         - &           - &          - &       - &           - &        - &            - &        - &            - &         - &             - \\
2022-08-31 &       MER & 9822.680 &   1800 &      7.0 &       4.0 &         7.0 &        3.0 &       - &           - &        - &            - &   -955.9 &         33.2 &         - &             - \\
   2022-11-09 &        IAC & 9892.758 &   1200 &     38.0 &      35.0 &        26.0 &       30.0 &   -51.5 &         1.5 &    -43.5 &          1.5 &  -1050.3 &          2.0 &   -202.34 &          1.43 \\
\hline
\end{longtable}

\section{Sinusoidal models of the \texorpdfstring{$\langle B_\textrm{z} \rangle$\ }\ Time Series} 
\label{sec:append_bz_fit}
\restartappendixnumbering

For each $\langle B_\textrm{z} \rangle$ dataset paired with each period, we use the Nelder-Mead method through the \texttt{scipy.optimize} package \citep{2020SciPy-NMeth} to find the optimal $A$, mean, and $\phi_0$ for each sinusoidal model. We show these parameters in Table \ref{tab:cos_fit} along with the extrema of the fits, Preston-$r$ value, and reduced $\chi^2$.  

\begin{table*}[!h]
\caption{\label{tab:cos_fit} Best parameters for the co-sinusoidal ($A\cos(\phi-\phi_0)+m$) fits to the $\langle B_\textrm{z} \rangle$ measurements of \citetalias{2021MNRAS.501.2677D} and \citetalias{2021MNRAS.501.4534J}, for the various periods listed in the text. We note that this fit could also correspond to an offset dipole. Additionally, we list the extrema of $\langle B_\textrm{z} \rangle$ predicted by the fit along with the associated $r$ value that constrains the relationship between $i$ and $\beta$ \citep{1967ApJ...150..547P}. We also list the reduced $\chi^2$. }

\begin{tabular}{rlrrrrrrr}
\toprule
$P$ (d) &    Dataset &     $A$ (kG) &  mean (kG) &   $\phi_0$ &   max &    min &      $r$ &  $\chi^2_\mathrm{red}$ \\
\hline
157.99 &     DU2021 & 2.863 & 3.054 &  0.627 & 5.917 &  0.191 &  0.032 &          10.264 \\
157.99 &  J2021 & 1.836 & 1.837 &  0.586 & 3.673 &  0.001 &  0.000 &           5.458 \\
153.17 &     DU2021 & 2.907 & 3.082 & -0.177 & 5.988 &  0.175 &  0.029 &          12.088 \\
153.17 &  J2021 & 1.818 & 1.766 & -0.056 & 3.583 & -0.052 & -0.015 &           5.233 \\
306.56 &     DU2021 & 4.655 & 0.460 & -0.162 & 5.116 & -4.195 & -0.820 &           1.104 \\
306.56 &  J2021 & 3.024 & 0.278 & -0.237 & 3.302 & -2.746 & -0.832 &           3.494 \\
\hline
\end{tabular}

\end{table*}

\bibliography{Seadrow_2025}
\bibliographystyle{aasjournal}

\end{document}